\documentclass[fleqn,usenatbib]{mnras}

\usepackage{newtxtext,newtxmath}

\usepackage[T1]{fontenc}

\DeclareRobustCommand{\VAN}[3]{#2}
\let\VANthebibliography\thebibliography
\def\thebibliography{\DeclareRobustCommand{\VAN}[3]{##3}\VANthebibliography}

\usepackage{graphicx}	
\usepackage{amsmath}	
\usepackage{multirow}
\usepackage{url}

\newcommand{\heii}{\mbox{He\,{\sc ii}} $\lambda$}
\newcommand{\hei}{\mbox{He\,{\sc i}} $\lambda$}

\newcommand{\niii}{\mbox{N\,{\sc iii}} $\lambda$}
\newcommand{\siiii}{\mbox{Si\,{\sc iii}} $\lambda$}

\newcommand{\ciii}{\mbox{C\,{\sc iii}} $\lambda$}
\newcommand{\civ}{\mbox{C\,{\sc iv}} $\lambda$}
\newcommand{\oiii}{\mbox{O\,{\sc iii}} $\lambda$}

\title[Fundamental parameters of the massive eclipsing binary system HM1 8]{Fundamental parameters of the massive eclipsing binary HM1~8\thanks{This paper includes data obtained with the 6.5 m Magellan Telescopes located at Las Campanas Observatory, Chile.
}}

\author[C. N. Rodr\'iguez et al.]{C. N. Rodr\'{\i}guez$^{1,2}$\thanks{E-mail: cnrodriguez@fcaglp.unlp.edu.ar},
G. A. Ferrero$^{1,2}$,
O. G. Benvenuto$^{1,2}\thanks{Member of  the Carrera del Investigador
 Cient\'{\i}fico, Comisi\'on de  Investigaciones Cient\'{\i}ficas de la  Provincia
 de Buenos Aires (CIC).}$,
R. Gamen$^{1,2}$,
N. I. Morrell$^{3}$,
\newauthor R. H. Barb\'a$^{4}$,
J. Arias$^{4}$ and
P. Massey$^{5,6}$ 
\\
$^{1}$Instituto de Astrof\'{\i}sica de La Plata, IALP, CCT-CONICET-UNLP, Argentina\\ 
$^{2}$Facultad de Ciencias Astron\'omicas y Geof\'{\i}sicas, Universidad Nacional de La Plata, Paseo del Bosque S/N, B1900FWA La Plata, Argentina.\\
$^{3}$Las Campanas Observatory, Carnegie Observatories, Casilla 601, La Serena, Chile.\\
$^{4}$Departamento de Astronom\'{\i}a, Universidad de La Serena, Cisternas 1200 Norte, La Serena, Chile.\\ 
$^{5}$Lowell Observatory, 1400 W Mars Hill Road, Flagstaff, AZ 86001, USA.\\
$^{6}$Department of Astronomy and Planetary Science, Northern Arizona University, Flagstaff, AZ 86001-6010, USA.\\
}

\date{Accepted XXX. Received YYY; in original form ZZZ}

\pubyear{2021}

\begin{document}
\label{firstpage}
\pagerange{\pageref{firstpage}--\pageref{lastpage}}
\maketitle

\begin{abstract}
We present a comprehensive study of the massive binary system HM1~8, based on multi-epoch high resolution spectroscopy, $V$-band photometry and archival X-ray data.
Spectra from the {\it OWN Survey}, a high resolution optical monitoring of Southern O and WN stars, are used to analyse the spectral morphology and perform quantitative spectroscopic analysis of both stellar components. The primary and secondary components are classified as O4.5~IV(f) and O9.7~V, respectively. 
From a radial-velocity (RV) study we derived a set of orbital parameters for the system.
We found an eccentric orbit ($e=0.14 \pm 0.01$) with a period of $P = 5.87820 \pm 0.00008$~days. 
Through the simultaneous analysis of the RVs and the $V$-band light curve we derived an orbital inclination of $70.0^{\circ} \pm 2.0$ and stellar masses of $M_a=33.6^{+1.4}_{-1.2}~\text{M}_{\sun}$ for the primary, and $M_b=17.7^{+0.5}_{-0.7}~\text{M}_{\sun}$ for the secondary. 
The components show projected rotational velocities $v_1\sin{i}=105 \pm 14~\text{km~s}^{-1}$ and $v_2\sin{i}=82 \pm 15~\text{km~s}^{-1}$, respectively. 
A tidal evolution analysis is also performed and found to be in agreement with the orbital characteristics. 
Finally, the available X-ray observations show no evidence of a colliding winds region, therefore the X-ray emission is attributed to stellar winds.
\end{abstract}




\section{Introduction}
Massive stars are those that end their lives as core-collapse supernovae, creating neutron stars or black holes. 
They play a key role in the astrophysics of their host galaxy due to their ionising UV fluxes, powerful winds and extreme stellar properties \citep{zinnecker2007}. 
Usually found in star forming regions, they are excellent tracers of recent star formation, as their input dominate the total luminosity of the environment \citep{massey2003}. 
They also contribute to the chemical evolution of their host galaxy 
by means of their stellar winds and supernovae explosions (\citealt{maeder1998, langer2012} and references therein). 
From a spectroscopic point of view, massive stars are found as main sequence O and early B types, and later as Wolf-Rayet (WR) stars and red supergiants, which evolve to their explosive end in a few million years or less.
In spite of their importance, our knowledge on some properties of these stars, such as their formation mechanisms and their masses, is still incomplete.

Massive stars show a high degree of multiplicity. 
In the Milky Way, at least 70\% of the O-type stars are found in binaries and multiple systems \citep{barba2017,sana2012}. 
Detailed studies of multiple stars allow derivation of important physical parameters, such as absolute masses and radii.
Different scenarios have been proposed for the formation of massive stars \citep{zinnecker2007}, although none of them entirely explains the precise forming mechanism; the orbital parameters of multiple systems can provide information on enviromental conditions during star formation \citep{mahy2013}. 
Although the evolution of massive stars is mainly determined by initial masses and mass-loss rates \citep{chiosi1986}, multiplicity may affect the evolutionary path of the components in short-period systems \citep{sana2010,langer2012,mahy2013}.

\begin{table*} 
	\centering
	\caption{Technical characteristics of the spectrographs used for this work.}
	\label{tab:spectrographs}
	\begin{tabular}{lllcc} 
		\hline
		Observatory & Telescope & Spectrograph & Spectral range    & Resolving power \\
		            &           &              &  [\AA] &            \\ 
		\hline
		Las Campanas & Ir\'en\'ee du Pont (2.5 m) & Echelle & $\sim 3600-10100$       & $\sim 20\,000$ \\
		La Silla     & MPG/ESO (2.2 m)            & FEROS   & $\sim 3600-9200$        & $\sim 48\,000$ \\
		\multirow{2}{3cm}{Las Campanas} & \multirow{2}{4cm}{Clay-Magellan (6.5 m)}    & \multirow{2}{0.5cm}{MIKE} & $\sim 3200-5000$ (blue) & $\sim 28\,000$ \\
		                                &                                           &                         & $\sim 4900-10000$ (red) & $\sim 22\,000$\\           
		\hline
	\end{tabular}
\end{table*}

Since the era of {\it Einstein} X-ray telescope, it is known that massive stars earlier than mid-B emit X-rays \citep{harnden1979}. 
Their origin is still debated but the most accepted models attribute this emission to plasma heated by shocks formed from intrinsic instabilities associated with radiatively-driven stellar winds (e.g. \citealt{lucy1980,owocki1988,feldmeier1997,owocki2013}). 
The X-ray luminosity is therefore expected to correlate with stellar parameters \citep{sciortino1990,gomezmoran2018}. 
In particular, it has been found that the X-ray luminosity $(L_{\rm X})$ of O-type stars linearly scales with their bolometric luminosities $(L_{\rm BOL})$ (\citealt{gomezmoran2018} and references therein). 
Then, if the ratio $L_{\rm X}/L_{\rm BOL}$ can be obtained from observations, it could be inferred where the X-ray radiation comes from (if the star is actually emitting it or a colliding-wind region exists), and help us constrain the astrophysical models.

In this work we concentrate on HM1~8 (HM1~VB10; $\alpha=~17^{\rm h}\,19^{\rm m}\,04^{\rm s}\!.4$, $\delta=-38^{\circ}\,49'\,05''$), a member of the open cluster Havlen \& Moffat No. 1 (HM1 = C~1715-387; $l=348.7^{\circ}$, $b=-0.8^{\circ}$). 
This is a compact group of stars situated in the inner part of our Galaxy, beyond the Sagittarius arm ($V_0-M_V=12.6$ or $d=3.3$~kpc), and highly reddened ($E_{B-V}=1.84\pm0.07$ mag) according to \citet{vazquez2001}.
HM1 8 was firstly classified as an O8 star by \citet{havlen1977}, and subsequently reclassified as O5\,V \citep{massey2001}, and O5\,III(f) \citep{gamen2008,sota2014}. 
Its binary nature was discovered by \citet{gamen2008} who found an orbital period of 5.9 days and minimum masses of 31~$\text{M}_\odot$ and 15~$\text{M}_\odot$, suggesting the secondary component is also an OB star. In the X-ray domain, \citet{naze2013} used data from the {\it XMM-Newton} satellite to study and characterize the properties of the stars in HM1, including HM1~8.

The goal of this work is to enlarge and deepen the study of HM1 8. 
The structure of the paper is as follows: the observations are described in Sec.~\ref{sec:obs}; sections~\ref{sec:spectral} and \ref{sec:analysis} present the optical spectroscopic and photometric studies of the system; an up-to-date X-ray study of HM1~8 is included in section~\ref{sec:rayosx}, and the tidal evolution of the pair is discussed in section~\ref{Sec:tidal}. 
Finally, results and some conclusions are summarized in Sec.~\ref{sec:conclusions}.


\section{Observations}
\label{sec:obs}
The data considered in this study comprise three categories: optical spectroscopy, optical photometry, and X-ray archival observations. Observations were carried out
at different facilities as explained in the following sections.

\subsection{Optical spectroscopy}
\label{sec:opt} 

The spectroscopic data used in this work were obtained under the OWN~Survey project between 2006 and 2014. 
This is a high resolution optical spectroscopic survey of a large sample ($\approx 300$ stars) of Southern O and WN stars for which no multiplicity indication was available at the beginning of the project \citep{gamen2008,barba2017}. 
In this case, the spectra were obtained with three spectrographs available at two different observatories in Chile: these are the \'echelle spectrograph attached to the 2.5 m du Pont telescope, the Magellan Inamori Kyocera Echelle (MIKE) at the 6 m Magellan Clay telescope,  both at Las Campanas Observatory, and the Fibre-fed optical echelle spectrograph (FEROS), at the 2.2 m telescope at La Silla Observatory. 
Table~\ref{tab:spectrographs} summarises the main properties of each instrumental configuration.
Spectra from Las Campanas Observatory were reduced using the standard  {\sc iraf}\footnote{{\sc iraf} was distributed by the National Optical Astronomy Observatory,
which was operated by the Associated Universities for Research in
Astronomy, Inc., under cooperative agreement with the National Science
Foundation.} routines; in the case of du Pont a $2\times2$ binning was applied to the CCD. FEROS spectra were reduced with the standard reduction pipeline in the MIDAS package provided by the European Southern Observatory (ESO).
Typical exposure times were from 30 to 40~minutes, which provided spectra with signal-to-noise ratios (SNR) ranging from 30 in the blue to 100 in the yellow-red region of the spectrum.

\subsection{Optical photometry}
\label{sec:optphot} 

Photometric observations were obtained during a monitoring of stars in OB associations looking for eclipsing massive binaries \citep{massey2012}. 
The observations used in this work were obtained in the $V$ filter, with exposure times of 10-30~s typically. 
The data were collected with two different telescopes: the 1.0 m Yale telescope at Cerro Tololo Interamerican Observatory, operated by SMARTS, and the 1 m Swope telescope at Las
Campanas Observatory. 
Table~\ref{tab:phot} summarises the principal characteristics of each instrumental setup. 
The data were reduced and measured by P.M. (SMARTS dataset) and R.G. (Swope dataset).
The reader is referred to \citet{massey2012} for details on the reduction procedure. The
combined dataset includes a total of 924 $V$ measurements, with typical errors of about 0.007 mag., which are presented in Table~\ref{tab:photdata}.

\begin{table*}
	\centering
	\caption{Characteristics of telescopes and instruments used for photometric observations of HM1 8.}
	\label{tab:phot}
	\begin{tabular}{lccccc} 
		\hline
		Observatory & Telescope & Camera & Scale               & FOV      & Median seeing\\
		            &           &        & [arcsec pix$^{-1}$] & [arcmin] & [arcsec]     \\
		\hline
		Las Campanas & Swope (1 m)       & SITe3  & 0.435 & $15 \times 23$ & 1.60\\
		Cerro Tololo & SMARTS Yale (1 m) & Y4KCam & 0.289 & $20 \times 20$ & 1.67\\
		\hline
	\end{tabular}
\end{table*}

\begin{table}
	\centering
	\caption{$V$-band photometric data of HM1 8 used in this work. The full table with the 924 measurements and errors is available online as supplementary material.}
	\label{tab:photdata}
	\begin{tabular}{lccc}
	    \hline
	    HJD$-2\,400\,000$     &   $V$    & $\sigma_V$   & Telescope \\
	    \hline
        53\,134.7247	&	12.526	&	0.001	&	Swope	\\
        53\,134.7353	&	12.533	&	0.003	&	Swope	\\
        53\,134.7382	&	12.513	&	0.011	&	Swope	\\

        ...         &   ...     & ... &  ...     \\
        \hline
	\end{tabular}
\end{table}


\subsection{X-ray data}
\label{sec:xray}

HM1 was observed on 2010 March 10 by the {\it XMM-Newton} X-ray satellite during the revolution 1877 (Obs-Id. 0600080101) for 25~ks. 
The observation, centred on ($\alpha, \delta\text{)}_{J2000}$= ($17^{\rm h}~19^{\rm m}~00.^{\!\rm s}48$, $-38^{\circ}~48^{'\!}~00.^{\!\!''}\!5$), was acquired by the European Photon Imaging Camera (EPIC).
This instrument consists of three detectors, two Metal Oxide Semiconductor (MOS) cameras \citep{turner2001}, and one pn camera\footnote{The pn camera is a fully depleted, back illuminated detector with a $p^+$ back diode and a $n^+$ anode.} \citep{struder2000} operating in the 0.2-15~KeV range. 
The data were taken with the medium filter in full-frame mode.
The reduction and analysis were carried out with the XMM Science Analysis System ({\sc sas}) version 15.0.0, 
following the threads recommended by the {\it XMM-Newton} team\footnote{Available at \url{https://www.cosmos.esa.int/web/xmm-newton/sas-threads}.}. 
No background flare affected the observations, and no source is bright enough to suffer from pile-up. 
The latest calibrations were applied with the {\sc emmproc} and {\sc epproc} tasks. 
The events were filtered to retain only the patterns and photon energies likely for X-ray events: patterns 0-4 and energies 0.5 to 15 KeV for pn, and patterns 0-12 and energies from 0.5 to 10~KeV for MOS1/2 instruments.
After this, we obtained clean event lists (calibrated and filtered) which will be used in the subsequent analysis.


\section{Spectral analysis}
\label{sec:spectral}

\subsection{Radial velocity measurements}

The high-resolution spectrum of HM1 8 reveals the double-lined spectroscopic nature of the system. Given the difference in the spectral types of the stellar components, not all of the absorption lines show contribution from both stars.
To obtain the radial velocity (RV) measurements for the binary components, we adjusted a Gaussian function to the cores of selected spectral lines using the \texttt{splot} task of {\sc iraf}.
The measured lines were \hei5876, \heii5412, \heii4542, \ciii5696, \civ5812 for the primary, and \hei5876 for the secondary.
In the case of \hei5876, we used two Gaussian functions to fit the profiles of both components simultaneously. 
The remaining lines were fitted using a single Gaussian function.
The individual heliocentric RV measurements are listed in Table~\ref{tab:RV}.
Air wavelengths were taken from NIST Atomic Spectra Database Lines Form \citep{NIST_ASD}, for \hei5876, \ciii5696 and \civ5812 lines, and from \citet{striganov1968}, for the He \textsc{ii} lines.

\begin{table*}
	\centering
	\caption{Radial velocity measurements (in km $\text{s}^{-1}$) for the stellar components of HM1~8. The label ``a'' and ``b'' indicates primary and secondary components, respectively.
    RV values determined near conjuntions (marked with $\ast$) were not used in the orbital solution.}
	\label{tab:RV}
	\begin{tabular}{lrcccccccc} 
		\hline
		HJD & \multicolumn{2}{c}{\hei 5876} & \heii 5412 & \heii 4542 & \ciii 5696 & \civ 5812 & \multicolumn{2}{c}  Cross-correlation & Instrument\\
	2400000+ & \multicolumn{2}{c}{5875.62 \AA} & 5411.52 \AA & 4541.59 \AA & 5695.92 \AA & 5811.98 \AA & \\
	    & \multicolumn{1}{c}{a} & \multicolumn{1}{c}{b} & \multicolumn{1}{c}{a} & \multicolumn{1}{c}{a} & \multicolumn{1}{c}{a} & \multicolumn{1}{c}{a} &
	    \multicolumn{1}{c}{a} & \multicolumn{1}{c}{b} &
	    \multicolumn{1}{c}{ }\\
	\hline
53873.858       &93.5   &-227.0 &107.0  &65.4   &89.3   &95.4   &90.8   &-213.4 &du Pont echelle\\
53874.870       &109.4  &-255.4 &121.4  &154.0  &102.5  &113.0  &106.6  &-245.1 &du Pont echelle\\
53877.785       &-100.5 &-      &-99.7  &-77.5  &-109.2 &-108.4 &-115.1 &172.8  &du Pont echelle\\
53920.751       &78.7   &-209.0 &91.1   &90.3   &79.4   &77.1   &73.6   &-203.0 &du Pont echelle\\
53921.734       &116.6  &-275.3 &137.2  &126.2  &122.7  &112.8  &112.7  &-262.7 &du Pont echelle\\
53937.635       &-17.5  &-      &10.5   &13.4   &13.6   &-5.6   &$-2.5^\ast$      &$-41.6^\ast$      &du Pont echelle\\
53938.710       &100.3  &-264.0 &111.9  &117.8  &105.1  &108.9  &97.6   &-247.0 &du Pont echelle\\
53954.570       &-48.7  &-      &-51.7  &-54.4  &-76.4  &-69.1  &-75.1  &100.4  &MIKE\\
53954.586       &-41.6  &-      &-45.7  &-46.5  &-63.2  &-61.9  &-72.0  &88.7   &MIKE\\
53955.588       &22.4   &-      &-      &-      &47.0   &66.2   &$25^\ast$      &-      &MIKE\\
53987.508       &-36.6  &-      &-39.2  &-25.8  &-67.4  &-61.0  &$-17^\ast$     &-      &du Pont echelle\\
53988.504       &-157.6 &264.9  &-157.3 &-150.0 &-158.0 &-170.9 &-167.4 &282.0  &du Pont echelle\\
53989.517       &-95.4  &159.7  &-81.4  &-86.0  &-92.3  &-99.8  &-104.5 &174.9  &du Pont echelle\\
53990.520       &-13.3  &-      &17.1   &51.3   &20.8   &-1.8   &$-0.7^\ast$      &$-50.7^\ast$      &du Pont echelle\\
53991.527       &110.6  &-242.0 &101.3  &107.3  &148.5  &158.4  &99.6   &-236.3 &du Pont echelle\\
54198.868       &-18.5  &-      &3.5    &23.9   &-17.7  &-35.3  &$-2.5^\ast$      &$-58.4^\ast$      &du Pont echelle\\
54257.750       &-11.9  &-      &-10.1  &-      &-13.0  &-32.3  &$-7.1^\ast$      &$-51.4^\ast$      &du Pont echelle\\
54258.805       &-172.7 &267.1  &-159.8 &-149.7 &-172.1 &-165.1 &-163.7 &263.8  &du Pont echelle\\
54600.726       &-123.5 &176.0  &-110.5 &-111.6 &-121.1 &-122.8 &-129.6 &183.4  &FEROS\\
54626.784       &120.4  &-282.7 &141.6  &115.5  &110.5  &127.4  &112.4  &-261.6 &FEROS\\
54953.803       &-73.0  &-      &-65.0  &-63.4  &-78.9  &-79.4  &-88.4  &108.6  &FEROS\\
54956.807       &80.1   &-203.6 &80.7   &85.8   &73.2   &79.9   &74.8   &-194.0 &FEROS\\
54961.793       &113.4  &-293.8 &131.0  &142.8  &107.0  &117.0  &105.9  &-272.1 &du Pont echelle\\
54964.682       &-164.7 &273.1  &-138.2 &-147.3 &-153.5 &-165.1 &-179.0 &271.0  &du Pont echelle\\
54976.747       &-139.8 &207.1  &-126.0 &-131.4 &-135.3 &-135.7 &-149.0 &218.9  &FEROS\\
56813.780       &115.4  &-261.6 &143.1  &117.2  &118.4  &120.3  &109.8  &-251.7 &du Pont echelle\\
56815.818       &-150.4 &253.7  &129.3  &-135.3 &-152.0 &-156.7 &-158.7 &262.2  &du Pont echelle\\
	\hline
	\end{tabular}
\end{table*}

\subsection{Spectral disentangling}

\begin{figure*}
	\includegraphics[width=\linewidth]{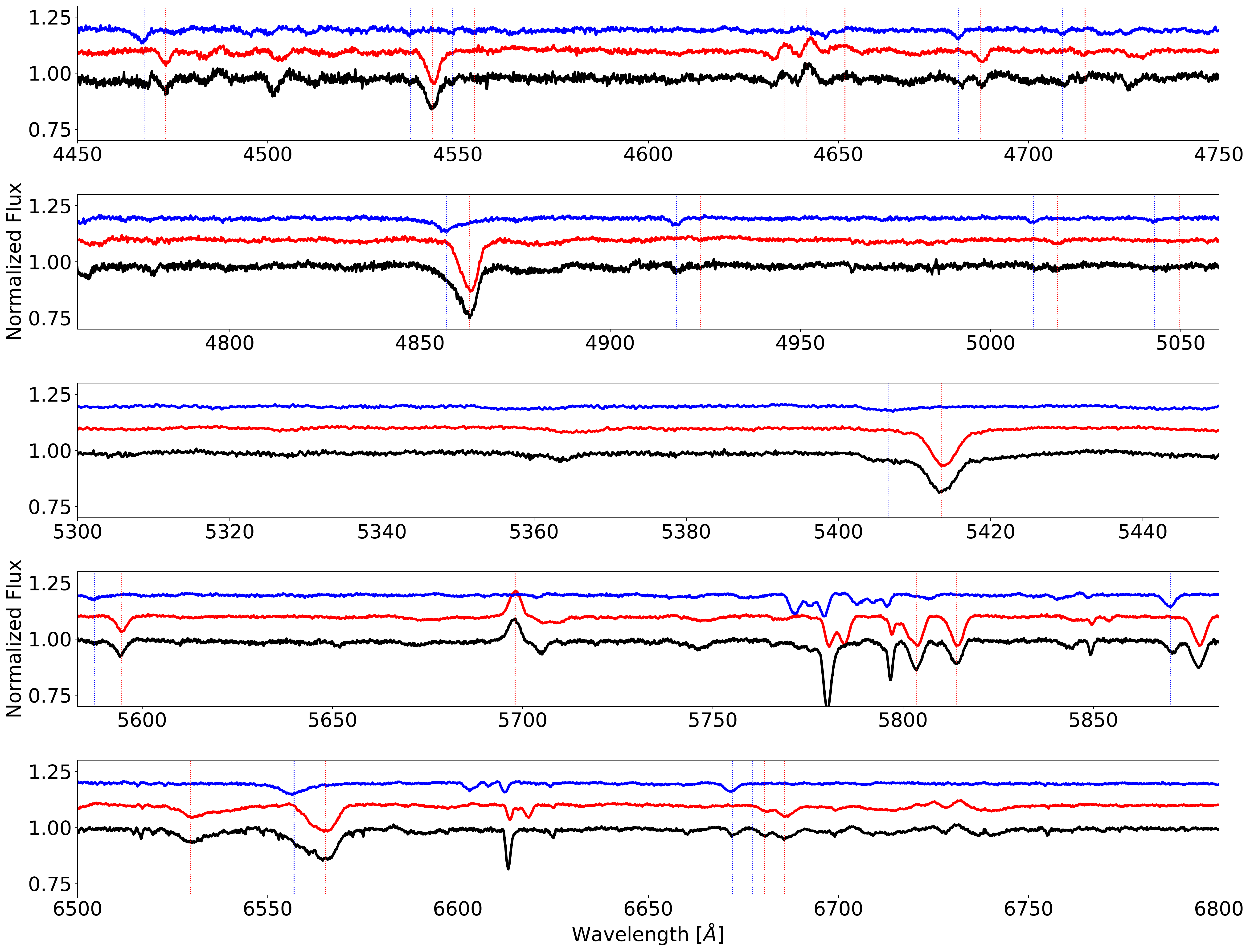}
    \caption{Templates obtained from the disentangling method (primary in red and secondary in blue), compared with a composite spectrum generated by combining the best four spectra during the upper quadrature of the primary star. 
    In each panel, we draw vertical lines depicting the (shifted) wavelengths for the primary (in red) and secondary (in blue) components of most important spectral features: \hei4471, \heii4542, \siiii4553, \niii4634/41, \ciii4651, \heii4686, \hei4713, H$_\beta$, \hei4922, \hei5016, \hei5047, \heii5412, \oiii5592, \ciii5696, \civ5801-12, \hei5876, \heii6527, H$_\alpha$, \hei6678, and \heii6683.
    }
    \label{fig:atlas-sec}
\end{figure*}

The RV measurements above described provide a starting point to apply a spectral disentangling method in order to obtain the individual spectra of each system's component.
In this case, we implemented the method developed by \citet{gonzalez2006}, following the procedure applied by \citet{2020MNRAS.494.3937B} in the analysis of the O+O system HD\,54662\,AB.
Compared to HD 54662 AB, HM1 8 has the advantage of  presenting orbital phases for which the spectral lines of the components are clearly separated, and the profiles are not blended at all (e.g. \hei5876).
To start the process we used {\sc fastwind} models as initial templates, corresponding to $T_{\rm eff}$ of 41\,000 K and 33\,000, for the primary and secondary, respectively,  and $\log g = 4.0$ for both components. 
Initial values for the projected rotational velocities ($v \sin i$) were estimated from the Gaussian fitting of the corresponding \hei5876 profiles. 
The methodology used to obtain {\sc fastwind} models is presented in Section \ref{sec:quant_analysis}.
The disentangling process converged very well after about 20 iterations. 
The convergence was greatly improved by centering the cross-correlation process in a few strong lines: \heii4542, \heii5412, \oiii5592, \civ5812, \hei5876 and \heii6683 for the primary, and \hei4922, \hei5876 and \hei6678 for the secondary component. 
The final disentangled spectra for both components of the system (component A and component B templates) are obtained from the combination of ten spectra with good SNR obtained at or near the orbital quadratures.
Both templates are shown in Fig~\ref{fig:atlas-sec}, along with a composite spectrum obtained as the sum of four spectra at the upper quadrature of the primary component.
The resulting individual spectra were used to determine the RVs of spectra not included in the disentangling process, applying a similar iterative method:
RVs of each component are determined iteratively via cross-correlation (using {\sc iraf fxcor}) after subtraction of the template of the companion star shifted to the appropriate RV. 
The convergence is reached after a few iterations.
The final RV measurements and their errors are listed in Table~\ref{tab:RV}. 
The comparison of these values with the RV measurements obtained using Gaussian fitting shows very consistent results and the RVs obtained through cross-correlation delivered the best spectroscopic orbital solution (see Section~\ref{sec:spec_orbit}).

Two characteristics of the disentangled spectra which arise in the large reddening that affects this star must be mentioned. First, the SNR is not uniform along the covered spectral range. 
For example the SNR in the blue ($\lambda4600$ \AA) is about 30 for most spectra; thus, as the secondary contributes with about 20\% of the total light (see Section~\ref{sec:quant_analysis}) the SNR of its spectra is lower than 10. 
With this situation, small differences in the continuum normalization could affect the disentangling result. 
In the next subsection we compare the template B with standard stars for spectroscopic classification as a proof of the reliability of our methodology.
A second characteristic is the large number of strong and broad diffuse interstellar bands (DIBs) distributed along the spectrum, mostly in the yellow-red portion.
These DIBs were not removed before the disentangling process, producing a wobbly appeareance of the continuum or structured profiles in some DIBs (e.g. $\lambda 5870$). 
Fortunately, these DIBs did not affect the main lines used for spectral classification and RV determinations.

\subsection{Spectral classification}

The spectral classification of both components was performed based on the templates obtained from the disentangling method, verifying that all the spectral lines were present in the original composite spectrum in quadrature. 
We followed the schema presented by \citet{sota2011} \citep[updated in ][]{2016ApJS..224....4M} in {\em The Galactic O-star Spectroscopic Survey} (GOSSS).
The template spectra were resampled to mimic the resolving power of GOSSS spectra, i.e. $R \approx 2500$.

\begin{figure*}
	\includegraphics[width=\linewidth]{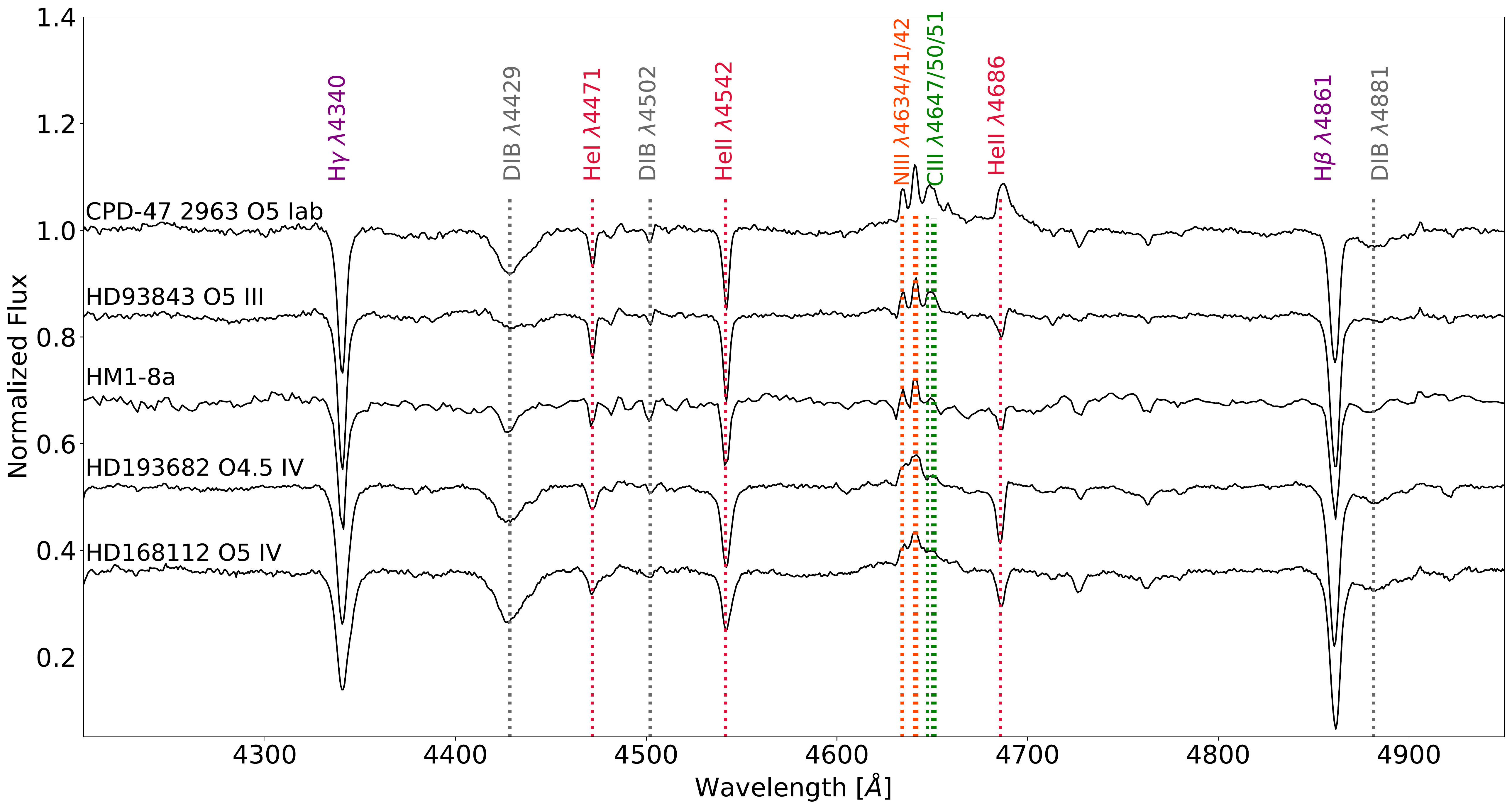}
    \caption{Comparison of the spectrum of HM1~8a 
    with the standards defined by \citet{2016ApJS..224....4M}.}  
    \label{fig:atlas-prim}
\end{figure*}

The spectrum of the primary component presents the \hei4471 line considerably weaker than the \heii4542 line.
A detailed comparison with the standard stars HD~93843 and HD~193682, corresponding to the types O5\,III and O4.5\,IV, respectively, shows that the primary is closer to HD~193682 (see Figure~\ref{fig:atlas-prim}).
Additionally, the strength of the \niii4634/41/42 emissions and the almost filled \heii4686 absorption indicate the (f) qualifier. 
Thus, we classified the primary star of HM1 8 as O4.5\,IV(f)\footnote{It is worth to note that the profile of \heii4686 is similar to that of HD\,93843, suggesting a luminosity class III. However, as the O4\,III spectral class does not have an assigned standard yet, we do not extrapolate following rigorously the Morgan-Keenan (MK) spectral classification process.}.

\begin{figure*}
\centering
    \includegraphics[width=\linewidth]{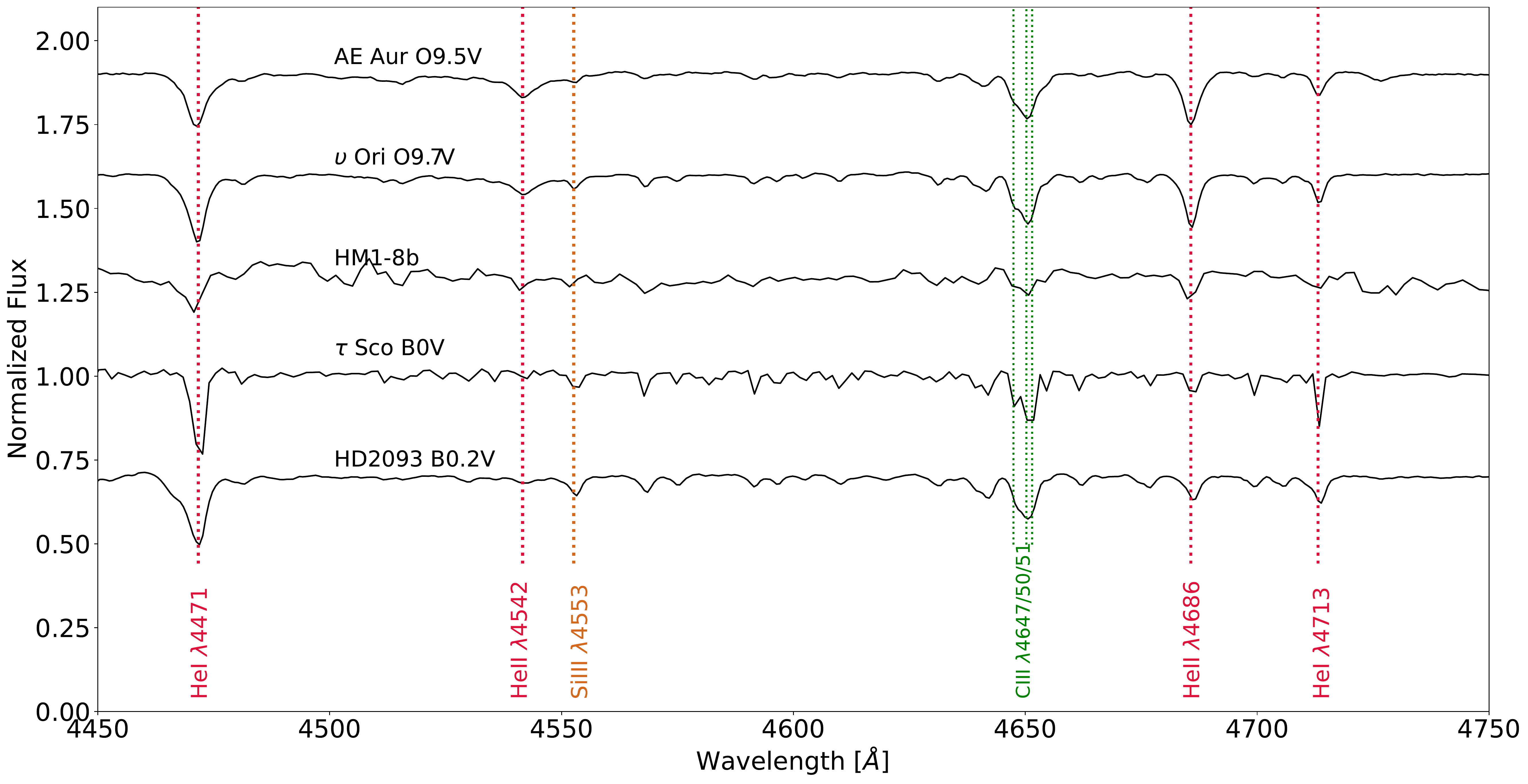}
    \caption{Comparison between the template of the secondary component (HM1~8b) and the spectroscopic standards for O9.5\,V, O9.7\,V, B0\,V, and B0.2\,V spectral types.
    }
    \label{standardb}
\end{figure*}

In the spectrum of the secondary component of HM1 8, we identified  He\,{\sc i}  $\lambda\lambda$4388, 4471, and 4713, \heii4686,  weak \heii4542 and \siiii4553 absorption lines, indicating a late-O or early-B spectral type.
Besides, the relative strengths of \heii4686 and \hei4713 point to luminosity class V.
In Fig.~\ref{standardb}, we plot the spectrum of HM1 8b along with the GOSSS \citep[][]{2016ApJS..224....4M} classification standards AE~Aur (O9.5~V), $\upsilon$ Ori (O9.7~V); and the \citet[][]{sota2011} standards $\tau$ Sco (B0~V) and HD~2083 (B0.2~V).
Based on the  strength of \heii4542 compared to \siiii4553, it can be seen that HM1 8b more closely resembles $\upsilon$\,Ori than  $\tau $\,Sco, a conclusion also supported by the relative strengths of \heii4686 and  \hei4471.
Thus, we classified this component as O9.7\,V.

\subsection{Spectral line-broadening analysis}
\label{sec:quantitative}

A spectral line-broadening analysis of the spectra was performed in order to study the rotational velocity of each star, as well as to identify other possible macro-broadening mechanisms. 
We used the {\sc iacob-broad} tool \citep{simondiaz2014}, which characterises the line-broadening of OB-type stars using a combined Fourier Transform (FT) and Goodness-Of-Fit (GOF) methodology. 
{\sc iacob-broad} allows the user to estimate the projected rotational velocity ($v sin\,i$) and the macroturbulence broadening ($v_{\rm mac}$) of the stars from the high resolution spectra.
According to \cite{simondiaz2007}, metallic absorption lines are more suitable for this analysis since they are less affected by other broadening effects besides rotation. 
Hence, we considered the \civ5812 line for the primary component, but for the secondary there was no metal line appropriate for this analysis, for what we considered \hei5875.

The analysis was performed using both the disentangled templates, a composite of 3 observed spectra and a FEROS spectrum at the orbital quadrature. 
As we are working with the templates and an observed composite spectrum, and the tool contemplates individual stars, we needed to correct it by their respective dilution factors. 
We estimated them from their calibrated magnitudes (according to their spectral types), understanding the dilution factor as an estimation of the fraction of observed flux from the spectra that corresponds to a specific component (we estimated dilution factors of $0.8$ and $0.2$ for the primary and secondary component, respectively).
As will be shown below, these dilution factors permit comparison of the templates with the determined models accordingly.

\begin{figure}
    \includegraphics[width=\linewidth]{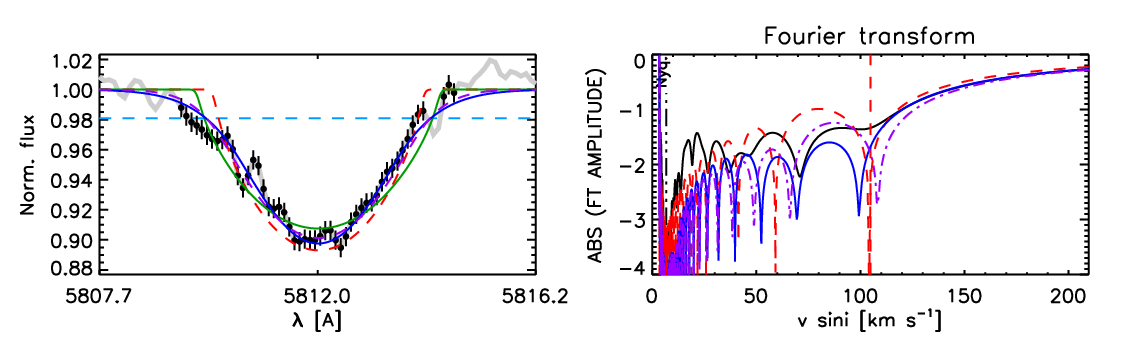}
     \includegraphics[width=\linewidth]{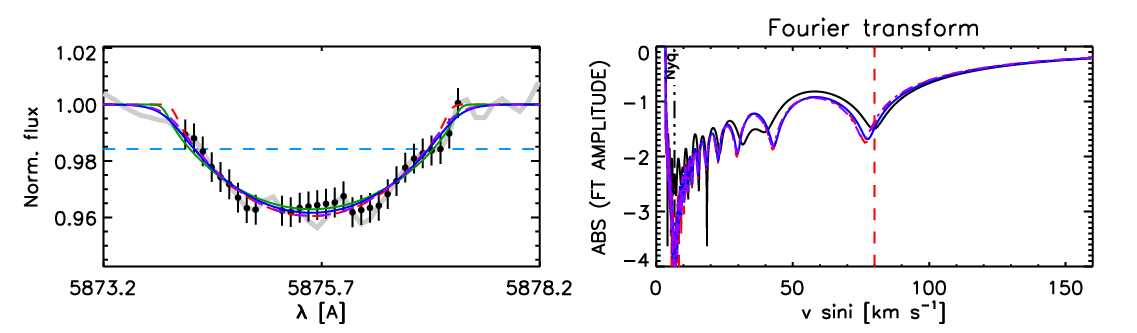}
    \caption{Characterisation of line-broadening of HM1~8 primary (top) and secondary (bottom) components obtained with the {\sc iacob-broad} tool.
    Left: \civ5812 of the primary and \hei5875 for the secondary in the composite spectrum (black line) with superimposed: 
    Fourier Transform (FT) fit (red), 
    Goodness-Of-Fit (GOF) (blue),
    GOF without considering macroturbulence (green), 
    and GOF taking into account the projected rotational velocity from the FT (purple).
    Horizontal dashed light blue line: estimated level of noise.
    Right: FT for the different methods, showing the first minimum (vertical dashed red line), where the projected rotational velocity is calculated (colours as in left panel). }
    \label{broad-prim}
\end{figure}

The best fits obtained in this analysis were obtained with the observed composite spectrum and they are shown in Fig.~\ref{broad-prim}. 
For the primary star we found $v sin\,i=105 \pm 14~\text{km~s}^{-1}$ and $v_{\rm mac}=67 \pm 6~\text{km~s}^{-1}$. 
The macroturbulence velocity, which represents all causes of macro-broadening other than rotation, is significant for this component.
This points to the existence of extra broadening mechanisms of different physical origins, for example, cyclic surface motions initiated by turbulent pressure instabilities \citep{simondiaz2017}.
The derived values for the secondary are $v sin\,i=82 \pm 15~\text{km~s}^{-1}$ and $v_{\rm mac}=22 \pm 7~\text{km~s}^{-1}$,
meaning that the broadening is fully dominated by rotation.

\subsection{Quantitative analysis}
\label{sec:quant_analysis}

We carried out a quantitative analysis of the system's components to estimate their atmospheric parameters: effective temperature $T_{\rm eff}$, surface gravity $\log g$ and wind strength $Q$ (where $Q=\dot{M}/\left(v_{\infty}R\right)^{1.5}$, being $v_\infty$ the wind terminal velocity). 

We employed the {\sc iacob} Grid-Based Automatic Tool ({\sc iacob-gbat}), which is an {\sc idl} package that compares the observed spectrum with a large grid of FASTWIND models \citep[][]{puls2005,1997A&A...323..488S}, convolved with their corresponding $v \sin\,i$ and $v_{\rm mac}$, and selects the one with the best fit (by means of a $\chi^2$ algorithm). 
In this way, {\sc iacob-gbat} performs a quantitative spectroscopic analysis based on standard techniques for O stars, using optical H and He lines \citep{Sim_n_D_az_2011}.

\begin{figure*}
	\includegraphics[width=\linewidth]{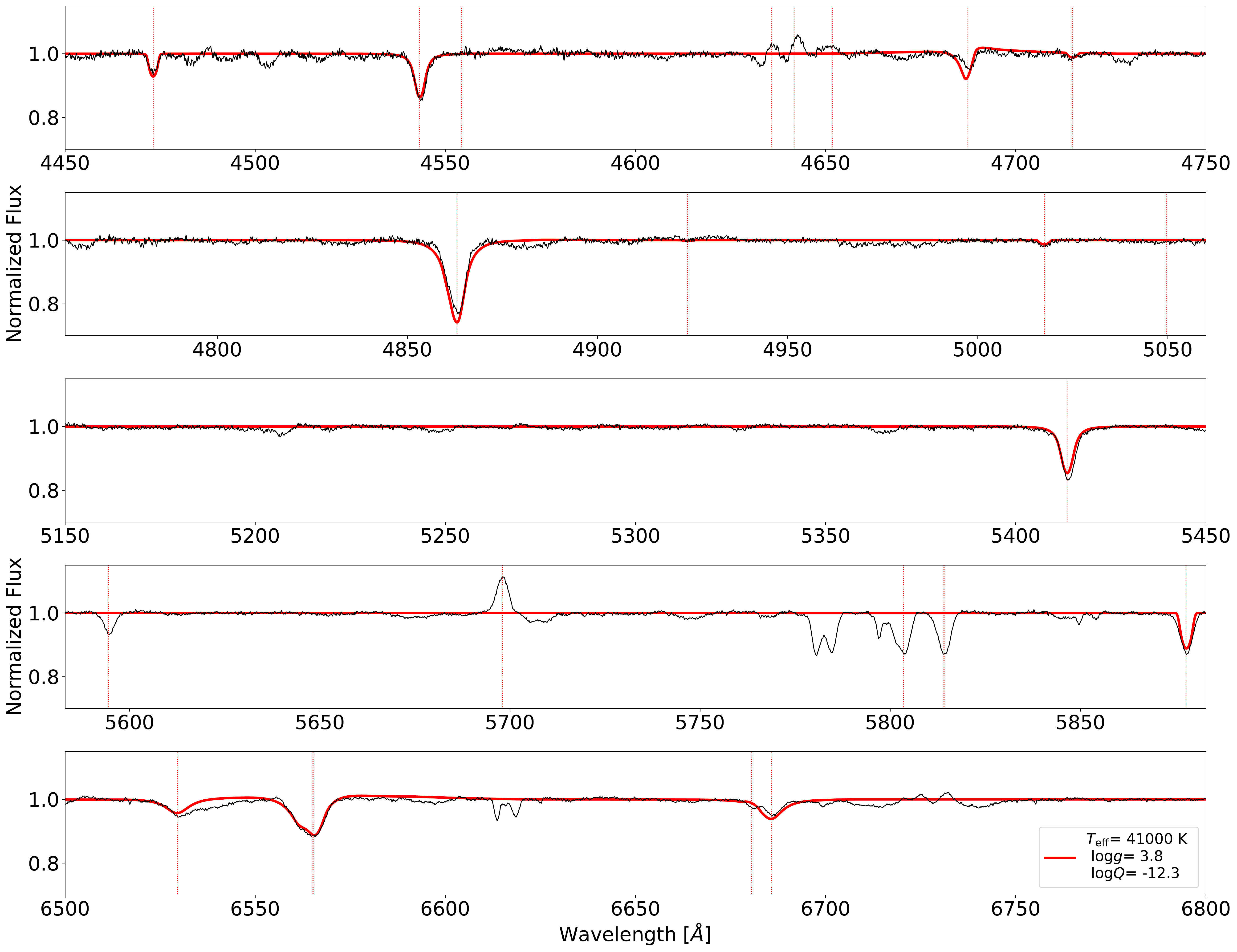}
    \caption{Template of HM1-8a obtained from the disentangling method, compared with a \textsc{fastwind} model (parameters in Table~\ref{tabla:iacob}). 
    In each panel, we draw vertical lines depicting the (shifted) wavelengths of the same spectral features indicated in Fig.~\ref{fig:atlas-sec}.
    }
    \label{fig:templatea}
\end{figure*}

We run {\sc iacob-gbat} on the template of HM1-8a using, as input parameters, $T_\mathrm{eff}$ and $\log{g}$ values for an O5~III star  \citep{martins2005}. 
We also adopted the broadening parameters found previously using the {\sc iacob-broad} tool, and fixing the associated helium abundances (Y$_\mathrm{He}$=0.10), the microturbulent velocities ($\xi_1$=5~km~s$^{-1}$), and the wind parameter $\beta$=0.8.
We fitted the following lines: H$\alpha$, H$\beta$, and H$\gamma$; \hei4471 and 5876; and \heii4542 and 5412.
The obtained parameters are shown in Table~\ref{tabla:iacob} and the comparison with the model is depicted in Fig.~\ref{fig:templatea}.
The overall agreement between the {\sc FASTWIND} model and the disentangled spectrum for the primary component is fairly good.
A wind component  ($\log{Q}=-12.3$) is needed in order to fit the \heii4686 and H$\alpha$ profiles.

\begin{figure*}
	\includegraphics[width=\linewidth]{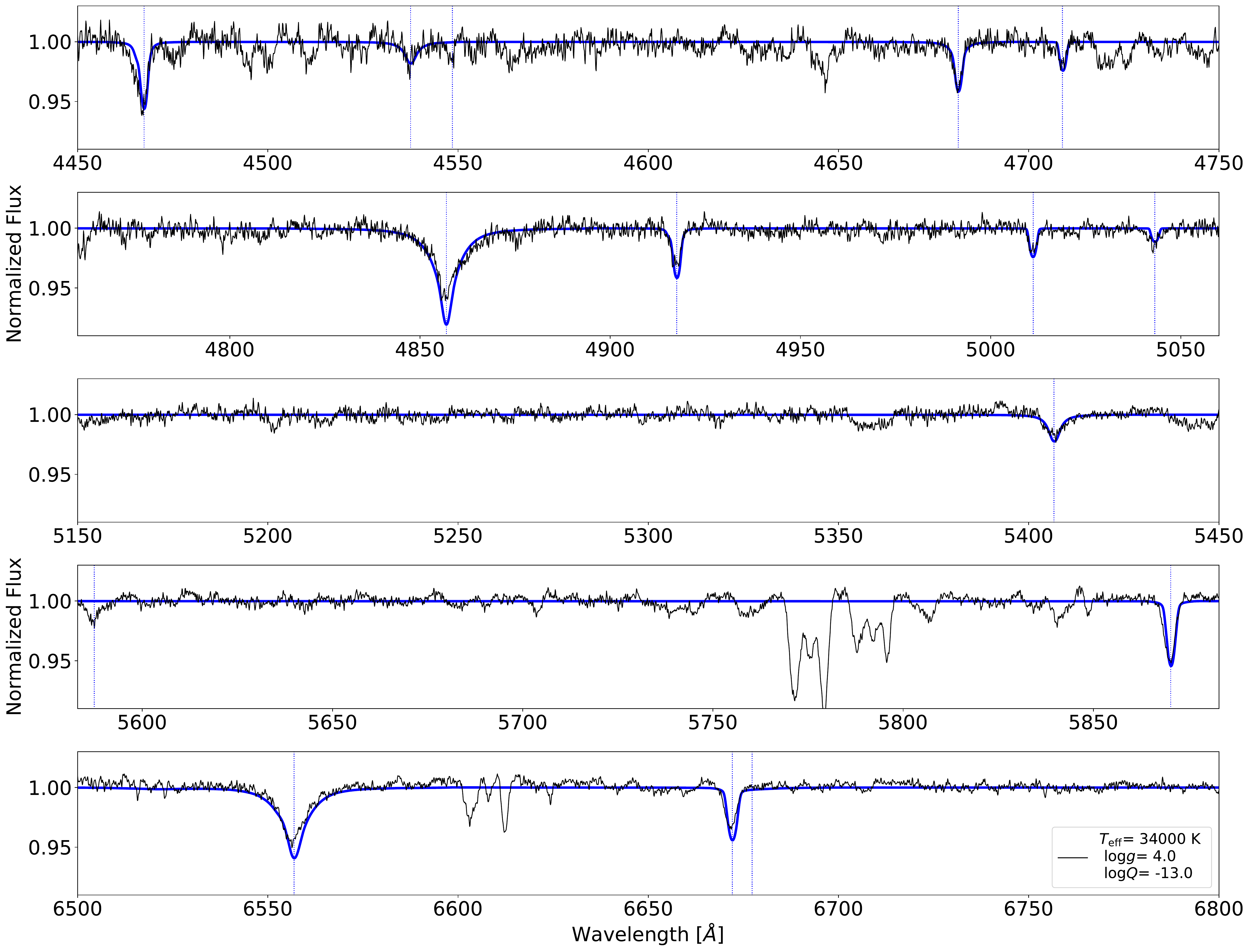}
    \caption{Template of HM1-8b obtained from the disentangling method, compared with a \textsc{fastwind} model (parameters in Table~\ref{tabla:iacob}). 
    In each panel, we draw vertical lines depicting the (shifted) wavelengths of \hei4471, \heii4542, \siiii4553, \heii4686, \hei4713, H$_\beta$, \hei4922, \hei5016, \hei5047, \heii5411, \oiii5592, \hei5875, \heii6527, H$_\alpha$, \hei6678, \heii6683.
    }
    \label{fig:templateb}
\end{figure*}

For HM1-8b, we also ran {\sc iacob-gbat} on its disentangled spectrum, providing as input parameters $T_\mathrm{eff}$ and $\log{g}$ values for an O9.5~V star \citep{martins2005}. 
The broadening parameters were adopted from the analysis with {\sc iacob-broad}. 
The fitted lines were: H$\alpha$, H$\beta$, H$\gamma$, \hei 4471, 4713, 4922, 5015 and 5876; and \heii4542, 4686 and 5411. 
The parameters were derived with $\log{g} = 4.0$ fixed, and are presented in Table~\ref{tabla:iacob}, while the comparison between the spectrum and the best FASTWIND model is illustrated in Fig.~\ref{fig:templateb}.
Again, the agreement between model and spectrum is fairly good.

An important parameter is the flux ratio between components, needed to calculate the dilution factors.
As we pointed previously, a flux ratio was estimated from the spectral types and magnitude calibration. 
To determine the dilution factors, we use the {\sc FASTWIND} models representing each component of the system (as they were calculated previously), and diluted them by several dilution factors from $f_1$=0.6-0.9 and $f_2$=0.1-0.4 for the primary and secondary, respectively. 
Then, we compared them with an observed spectrum and concluded that $f_1$=0.8 and $f_2$=0.2 are proper values.

\begin{table}
\caption{Parameters obtained from the quantitative spectroscopic analysis of both components of HM1-8.}
\centering                          
\begin{tabular}{c c c}       
\hline       \hline    \noalign{\smallskip}    
 Parameters &  Primary & Secondary \\
 \noalign{\smallskip}
\hline  \noalign{\smallskip} 
$ v \sin i$ [km~s$^{-1}$]     & 105 $\pm$ 14 & 82 $\pm$ 15 \\
$v_\mathrm{mac}$ [km~s$^{-1}$]&  67 $\pm$  6 & 22 $\pm$  7 \\
$T_\mathrm{eff}$ [K]& 41200 $\pm$ 1200 & 34500 $\pm$ 1200 \\
$\log{g}$ [dex] & 3.76 $\pm$ 0.15 &  4.0 (fixed) \\
$\log{Q}$ [dex] & $-12.3 \pm 0.1$ &  $-13.0 \pm 0.3$ \\
\hline  \noalign{\smallskip}
\multicolumn{3}{c}{Fundamental parameters adopting:}\\
$M_V$   [mag]  & -5.2 & -3.7 \\
           \hline  \noalign{\smallskip}
$R$ [R$_{\odot}$]          & 11.0 $\pm$ 0.2  & 5.7 $\pm$ 0.1\\
log ($L/L_{\odot}$)  [dex] & 5.49 $\pm$ 0.04 & 4.61 $\pm$ 0.04\\
$M_{sp}$ [M$_{\odot}$]     & 26.8 $\pm$ 8.2  & $< 9.7$ \\
\noalign{\smallskip} \hline 
\end{tabular}
\label{tabla:iacob}
\end{table}

\subsection{Struve-Sahade Effect} 
\label{sec:sseffect}

\begin{figure}
    \centering
	\includegraphics[width=\columnwidth]{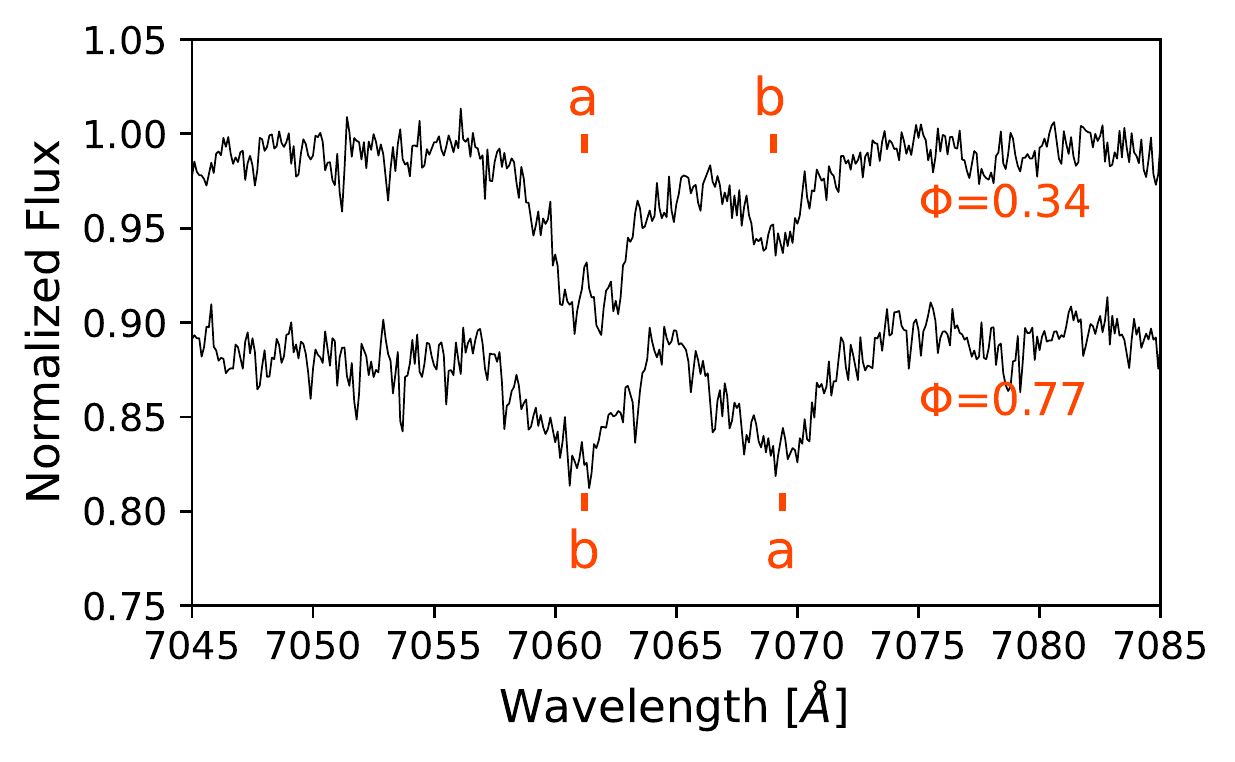}
	\includegraphics[width=\columnwidth]{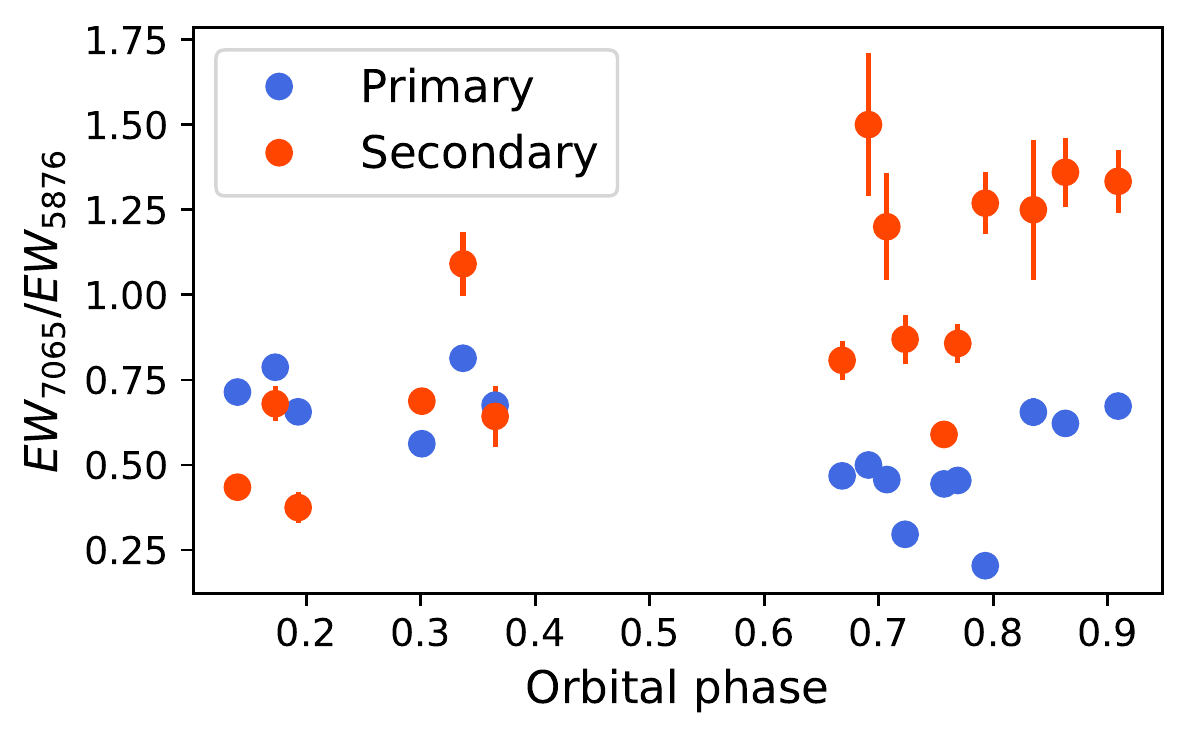}
    \caption{Top: two FEROS spectra around \hei7065, at different orbital phases (upper one with the secondary component receding and bottom one with the secondary approaching) where the S-S effect can be noticed.
    Bottom: \hei7065 to \hei5876 EW ratio vs orbital phase for both components of HM1~8. 
    It can be seen how the ratio change with the orbital phase, being more evident for the secondary component.}
   \label{fig:sseffect}
\end{figure}

During the visual inspection of the spectra, we noticed changes in the relative intensity of the components in the \hei7065 line at the two quadratures. 
These spectral variations could be indicative of the presence of the Struve-Sahade (S-S) effect.
We understand it as the apparent strengthening (weakening) of some absorption lines of both components when they are approaching (receding); which is different from the traditional concept where the lines associated with the secondary component were the ones that presented changes (see \citealt{linder2007}). 
Since its discovery \citep{bailey1896}, there were attempts to explain this effect without being possible to fully understand it, and with the years it became more likely the idea that there are several mechanisms producing this effect \citep{bagnuolo1999,linder2007,palate2013}. 
In the recent work of \citet{abdul2020}, they present a model that takes into account the 3-dimensional surface geometry of a system to produce spectral profiles at given phases and orientations, and they could represent this effect in the HD 165052 system. 
Fig.~\ref{fig:sseffect} shows the \hei7065 line in two FEROS spectra at two opposite quadratures to picture the S-S effect. 
Also, it is displayed the equivalent width (EW) ratio of \hei7065/\hei5876 lines (S-S effect is not observed in \hei5876), for both components.
To check the reliability of the spectral variations, we also measured the EW for the DIB $\lambda$5850 in the same spectra finding that it is constant at 5\% level, which is much smaller than the change in \hei7065, thus the equivalent width variations observed in this line are real and not due to noise or normalisation errors. 
It must be noted, however, that flanking the \hei7065 line there are three faint DIBs ($\lambda\lambda$7061.0, 7062.5 and 7069.0), which might slightly affect the EW measurements.

\section{The spectroscopic orbit}
\label{sec:spec_orbit}

As a first step, we verified that the new RV data are compatible with the periodicity reported by \citet{gamen2008}. 
This period was used as initial input, for the spectroscopic orbit calculation.
We then fitted the spectroscopic orbit to the different lines measured (see Table~\ref{tab:RV}) by means of the {\sc fotel} program  \citep[acronym for {\it Fotometric Elements}][]{hadrava2004}, deriving a RV orbital solution which can be combined with light-curve (LC) analysis of binary systems.
The code converged quickly to robust orbital solutions using the different RV datasets listed in Table~\ref{tab:RV}, letting free all the orbital parameters.
In particular, the orbital solutions obtained with \hei5876 and by means of the cross-correlation process using disentangled spectra shown the smallest errors, in special the latter one. 
Therefore, we consider that the spectroscopic orbit solutions obtained with the RVs determined by means the cross-correlation process as the most representative for this binary system. 
The parameters derived from this orbital solution are listed in the Table~\ref{tab:Fot_sol}.

\section{The eclipsing binary model}
\label{sec:analysis}

Phasing the photometric observations with the spectroscopic orbital period reveals a small and sharp drop in brightness at a phase close to the lower conjunction (the early-O star in the back side of the orbit). 
This feature can be understood as the detection of the primary eclipse of the system, i.e. when the secondary star is partially occulting the disk of the primary component, as depicted below in Fig.~\ref{fig:eclipse}.

\begin{figure*}
	\includegraphics[width=\linewidth]{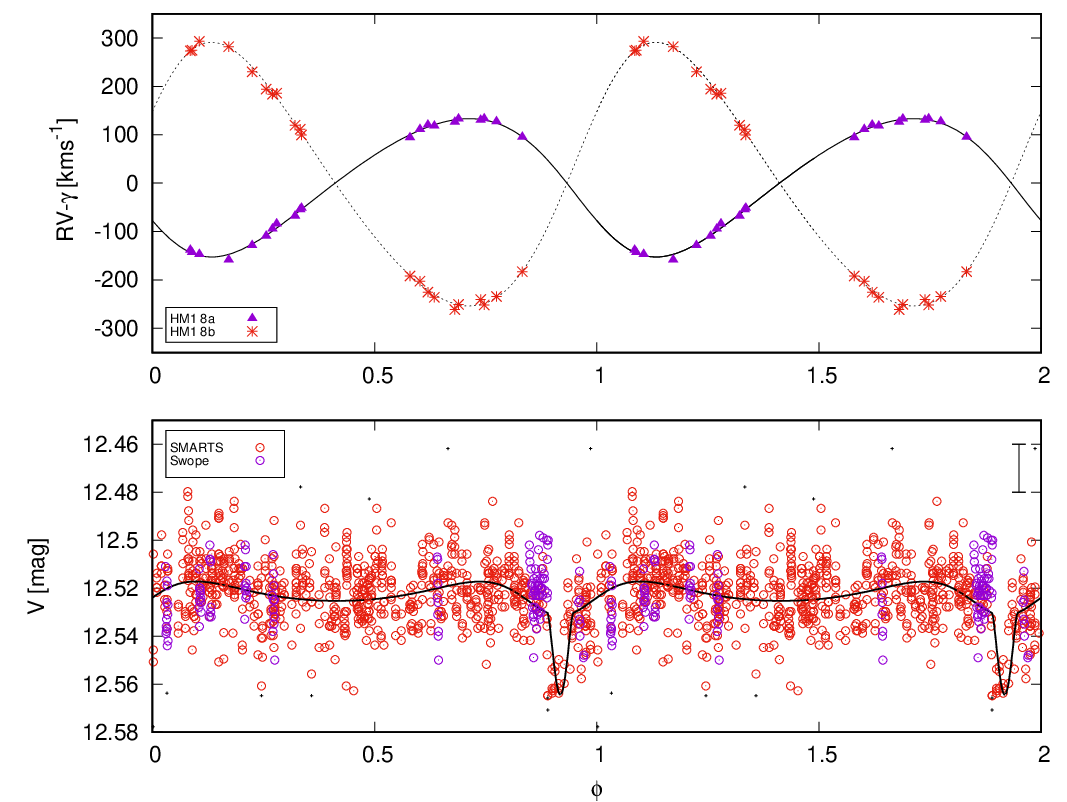}
    \caption{Upper panel: RV curve of HM1~8 obtained through cross-correlation method and phased using the ephemeris from Table \ref{tab:Fot_sol}.
    The systemic velocities determined for each component was subtracted. 
    Solid line and triangles: primary component; dotted line and crosses: secondary component.
    RV error bars are smaller than the symbols.
    Lower panel: light curve in $V$ filter, phased with the same period.  
    Red points: SMARTS Yale 1-m telescope; purple: Swope telescope. 
    Small dots: data not considered in the solution ($3 \sigma$ clipping).
    Solid line: {\sc fotel} model.
    Error bar represents a typical photometric error.
    }  
    \label{fig:mult}
\end{figure*}

Under the assumption of a sharp primary eclipse, we performed a modeling of the photometric time series by means of the {\sc fotel} code.
For this basic model, the user provides initial values for the orbital elements ($P$, $T_0$, $e$, $\omega$, $K_{1,2}$), and fixed values for the effective temperature of the stars, limb-darkening coefficients, magnitudes, and radii.
Orbital elements were fixed from the spectroscopic solution. 
Limb-darkening coefficients were taken from the \citet{castelli2003} calculations for a linear law, while $T_{\rm eff}$ and $\log{g}$ were adopted from the quantitative analysis performed with the \textsc{iacob-gbat} code.
Magnitudes in $V$-band for each component were calculated considering the dilution factors and the apparent magnitude of the system \citep[$m_V = 12.52$~mag;][]{reed2003}.
As the primary eclipse is grazing partial and no secondary one is detected, some constraints could be taken about radii.
The luminosity of each star can be obtained adopting a distance of $d=2805^{+146}_{-157}$~pc \citep{2021AJ....161..147B}, an extinction in the $V$-band, $A_V = 5.678 \pm 0.054$~ \citep{maizapellaniz2018}, bolometric corrections, $BC_\mathrm{a} = -3.67$~mag and $BC_\mathrm{b}=-2.97$~mag \citep[][]{martins2006}, and the calculated dilution factors (actually, the flux ratio).
Then, by means of the Stefan-Boltzmann formula, we can infer the radii.
Then, we ran \textsc{fotel} with three different set of radii (considering a range in luminosity) to determine the inclination of the system.
Derived parameters are listed in Table~\ref{tab:Fot_sol}, and the light curve model represented in Fig.~\ref{fig:mult} together to phased photometric observations.

Larger luminosity implies larger radius, and then, smaller orbital inclination in order to keep the depth of the eclipse, and therefore, larger absolute masses. 
Smaller masses are obtained in the case of smaller luminosities and radii.
In this way, we adopt as the error interval of the inclination these calculations made with different radii and luminosities, which is propagated as an estimate in the other orbital parameters that consider the inclination (semi-axes and masses).

In Fig.~\ref{fig:eclipse} we provide an illustration of the system's configuration at the inferior ($\Phi=0.9$) and superior ($\Phi=0.4$) conjunctions according to our orbital solution\footnote{This illustration was obtained with Phoebe-1.0 \citep{phoebe2005} by adopting the parameters in Table \ref{tab:Fot_sol}.}. 
Both stars are represented as seen from the observer.  
It can be noticed that due to the orbital eccentricity, a grazing primary eclipse is happening near the periastron passage, while the secondary one cannot occur due to larger separation of the stars in the superior conjunction.

\begin{table}
\centering
\caption{Solution given by {\sc fotel} putting together the RV measurements and the photometric data. 
}
\label{tab:Fot_sol}
\begin{tabular}{lccc} 
	\hline
	Element & \multicolumn{3}{c}{Value} \\
		\hline
	\multicolumn{4}{c}{Spectroscopic orbital solution} \\
		\hline
	$P$ [d]                      & \multicolumn{3}{c}{$5.87820\pm0.00008$} \\
	$T_{0}$ [HJD $-2\,400\,000$] & \multicolumn{3}{c}{$56\,815.30\pm0.09$} \\
	$e$                          & \multicolumn{3}{c}{$0.14\pm0.01$}\\
	$V_\gamma^\mathrm{a}$ [km $\text{s}^{-1}$]        & \multicolumn{3}{c}{$-20.2\pm1.6$} \\
	$V_\gamma^\mathrm{b}$ [km $\text{s}^{-1}$]        & \multicolumn{3}{c}{$-11.5\pm2.8$} \\
	$\omega \text{[}^{\circ}\text{]}$                 & \multicolumn{3}{c}{$119\pm10$} \\
	$K_\mathrm{a}$ [km $\text{s}^{-1}$]               & \multicolumn{3}{c}{$143\pm2$} \\
	$K_\mathrm{b}$ [km $\text{s}^{-1}$]               & \multicolumn{3}{c}{$273\pm3$} \\
	$q=M_{b}/M_{a}$                                   & \multicolumn{3}{c}{$0.52\pm0.02$} \\
	$a_\mathrm{a}\sin{i}$ [$\text{R}_{\odot}$]        & \multicolumn{3}{c}{$16.4\pm0.1$} \\
	$a_\mathrm{b}\sin{i}$ [$\text{R}_{\odot}$]        & \multicolumn{3}{c}{$31.3\pm0.4$} \\
	$M_\mathrm{a}\sin^{3}{i}$ [$\text{M}_\odot$]      &\multicolumn{3}{c}{ $28.4 \pm 0.7$ } \\
	$M_\mathrm{b}\sin^{3}{i}$ [$\text{M}_\odot$]      &\multicolumn{3}{c}{ $14.6 \pm 0.4$}  \\
	$\text{r.m.s}_{\text{(O-C)}^\mathrm{a}}$ [km s$^{-1}$] & \multicolumn{3}{c}{4.5} \\
	$\text{r.m.s}_{\text{(O-C)}^\mathrm{b}}$ [km s$^{-1}$] & \multicolumn{3}{c}{5.8} \\
\hline
\multicolumn{4}{c}{Fixed parameters} \\
\hline
	$V_\mathrm{a}$ [mag]                   & \multicolumn{3}{c}{$12.8$ }     \\
	$V_\mathrm{b}$ [mag]                   & \multicolumn{3}{c}{$14.3$  }   \\
	$T_{\rm eff}^\mathrm{a}$ [K]           & \multicolumn{3}{c}{ 41\,200 }\\
	$T_{\rm eff}^\mathrm{b}$ [K]           & \multicolumn{3}{c}{ 34\,500 }\\
	$M_\mathrm{Va}$ [mag]                  & -5.4 & -5.2  & -5.0\\
	$M_\mathrm{Vb}$ [mag]                  & -3.9 & -3.7  & -3.5\\
    $R_\mathrm{a}$ [$\text{R}_{\odot}$]    & 11.6 & 10.8  & 9.6\\
    $R_\mathrm{b}$ [$\text{R}_{\odot}$]    &  6.2 &  5.8  & 5.1\\		
\hline
\multicolumn{4}{c}{Derived parameters} \\
\hline
$i$ [$^{\circ}$]             &  68.0 & 70.0	& 72.0 \\
$a$ [$\text{R}_{\odot}$]     &  51.6 &	50.7 &	50.3 \\
$M_\mathrm{a}$ [$\text{M}_{\odot}$] &  35.0 &	33.6 &	32.4 \\
$M_\mathrm{b}$ [$\text{M}_{\odot}$] &  18.2 &	17.7 &	17.0 \\			$\text{r.m.s}_{\text{(O-C)~photometry}}$ [mag]           & \multicolumn{3}{c}{0.013} \\
		\hline
	\end{tabular}
\end{table}

\begin{figure}
	\includegraphics[width=\columnwidth]{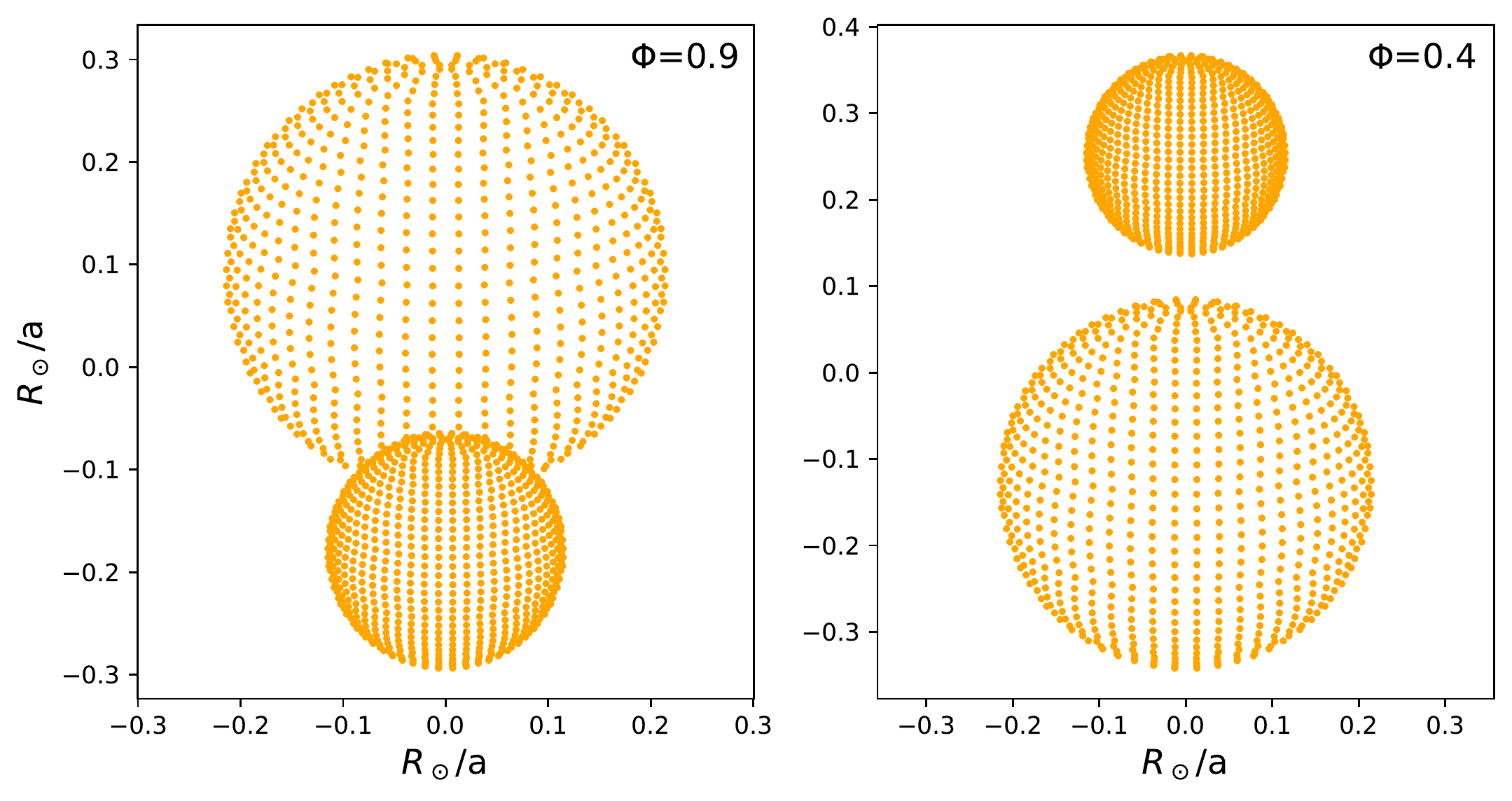}
    \caption{{\sc fotel} model configuration of the binary system HM1~8 projected in the plane of the sky during the conjunction phases.} 
    \label{fig:eclipse}
\end{figure}

\begin{figure}
	\includegraphics[width=1.\columnwidth]{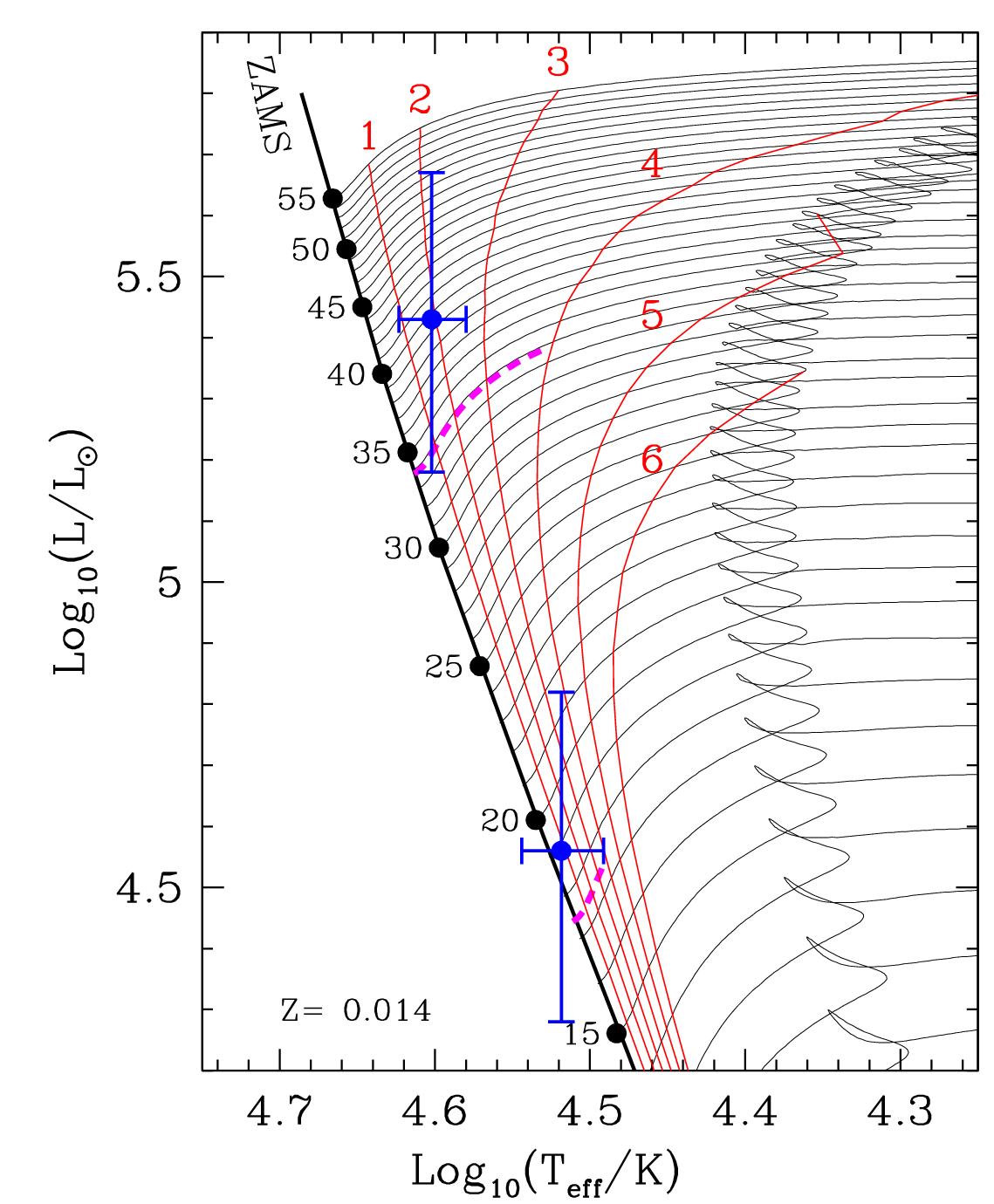} 
    \caption{$\log{L}-\log{T}_{\rm{eff}}$ diagram. 
    Black lines represent the evolutionary tracks for solar composition single stars from 11~${\rm M}_{\odot}$ to 55~${\rm M}_{\odot}$. Thin red lines denote isochrones every 1~Ma. 
    Thick dashed magenta lines indicate the evolution of stars with the masses presented in Table~\ref{tab:Fot_sol} up to an age of 4~Ma. 
    The error bars corresponding to the luminosity and effective temperature of each component, given in Table~\ref{tabla:iacob} are denoted in blue. 
    These evolutionary tracks were calculated employing the code described in \citet{2003MNRAS.342...50B}.}    
    \label{fig:hr}
\end{figure}

For completeness, we also plot the $\log{L}-\log{T_{\rm{eff}}}$ plane in Fig.~\ref{fig:hr} (where evolutionary tracks and isochrones were calculated with the code described in \citealt{2003MNRAS.342...50B}). We have assumed solar abundances with $Z=0.014$ and moderate overshooting with $\alpha_{\rm OV}=0.2$. 
In these calculations we have neglected the effects of stellar rotation on the evolution of the stars. 
This approach is fairly good considering that components in the pair rotate far slower than the breakup velocity (see below).
Taking into account the values for the masses of the components given in Table~\ref{tab:Fot_sol}, and the mass loss rates derived from the models, we deduce that their masses on the ZAMS were about $34^{+3}_{-2} {\rm M}_{\odot}$ and $17.9^{+2}_{-1} {\rm M}_{\odot}$, assuming an age of about 2\,Ma for the binary. 
Tracks for ages up to 4~Ma, corresponding to these masses are shown in Fig.~\ref{fig:hr} with thick magenta dashed lines. 
It is important to remark that this system will suffer a Roche lobe overflow at an age of 4.8~Ma. 
Clearly, the system is younger and the components of the pair are still detached.

\section{X-ray analysis}
\label{sec:rayosx}

In order to characterize the X-ray emission of HM1~8 system and to compare it with the results of \citet[][]{naze2013}, we performed a spectral analysis to the calibrated and filtered events lists from {\it XMM-Newton} observations (as explained in Sec.~\ref{sec:xray}), where we extracted and fitted a spectrum of each EPIC camera. 
In particular, we want to determine if the emission is associated with the winds of the stars or a colliding wind region (CWR). 

Firstly, we extracted a spectrum of HM1~8 for the EPIC MOS1/2 and pn cameras, following the spectrum extraction threads for EPIC cameras\footnote{\url{ https://www.cosmos.esa.int/web/xmm-newton/sas-threads}}, for which we acquired three different images obtained by filtering the event lists by three energy bands: soft (0.5-1.2~keV), medium (1.2-2.5~keV) and hard (2.5-10~keV), and combined them. 
We chose regions to extract the source+background and the background spectra to obtain the source spectrum.
The resultant spectrum was grouped to assure SNR=1 per bin.

\begin{figure}
	\includegraphics[width=\columnwidth]{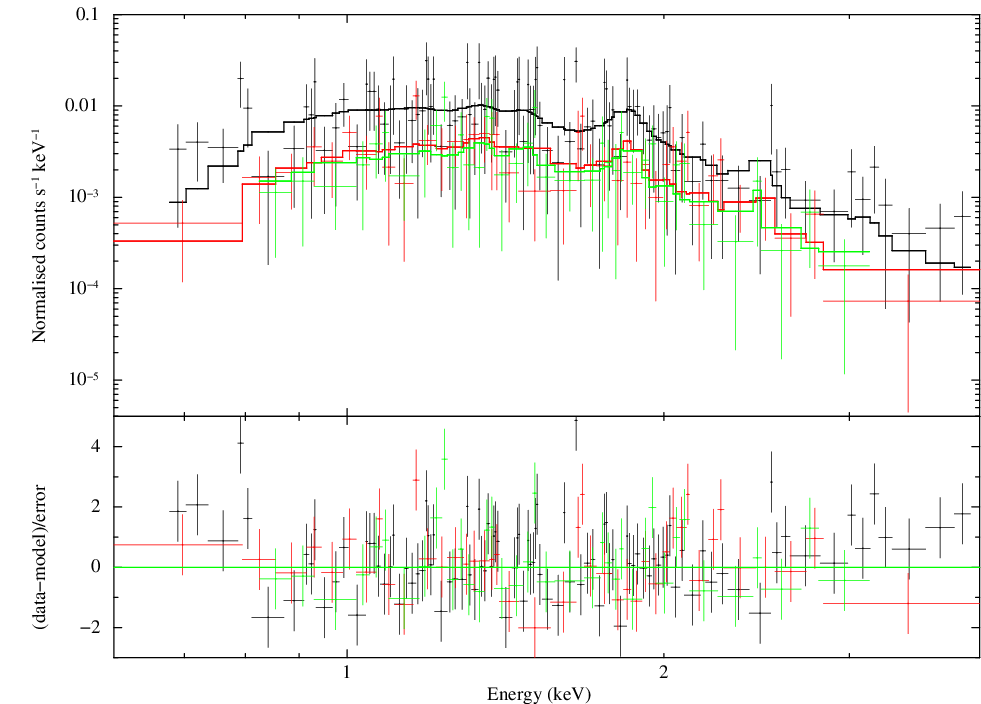}
    \caption{X-ray spectrum of HM1~8 for the three EPIC cameras with the fitted c*TBabs*APEC model. 
    Black: {\em pn} camera; red: MOS1 camera; green: MOS2 camera. The upper panel shows normalised counts vs. energy, and the lower panel represents the residuals.}
    \label{fig:x-hm18}
\end{figure}

Inspecting the images in the 3 energy bands we noticed that, for energies greater than 3~keV, there is no detectable signal from the source that could be distinguished from the background noise, so the spectral analysis was limited to the energy range 0.5-3~keV.
In this sense, we avoid possible noise contamination that could hinder the fit of the extracted spectrum. 
To fit the spectra, we used the {\sc xspec} package (version 12.11.1; \citealt{arnaud1996}), choosing an APEC model (an emission spectrum from collisionally-ionized diffuse gas, calculated with the AtomDB atomic database\footnote{\url{http://www.atomdb.org}}), modified by the Tuebingen-Boulder ISM absorption model (TBabs;  \citealt{wilms2000}), acting on the energy range 0.5-3~keV. 
Due to the low counts per channel in the spectrum, we used c-statistic to perform the fit; which assumes a Poisson distribution instead of a Gaussian one. 
Fig.~\ref{fig:x-hm18} shows the star spectrum for the three cameras and the best fit for each one, and Table~\ref{tab:model}  presents the fitted parameters. 
It can be seen that, considering the errors, our results are similar to those obtained by \citet{naze2013}.

\begin{table}
	\centering
	\caption{Parameters derived from the fitted with {\sc xspec} using model c*TBabs*apec. 
	\citet{naze2013} model: wabs*wabs*apec; where the first wabs was fixed with $N_{\rm H}=1.1 \times 10^{22}~{\rm cm}^{-2}$. In both works a solar abundance was adopted. The normalisation of the apec model is given as $10^{-14} \int n_e n_H dV / d^2$, where $d$ is the distance to the source (in cm), $n_e$ and $n_H$ are the electron and hydrogen densities of the source (in cm$^{-3}$). 
	Since counts of the {\em pn} camera are greater than the MOS cameras, we fixed c$_{\rm pn}=1.0$ and obtained the others with respect to this.}
	\label{tab:model}
	\begin{tabular}{lcc} 
		\hline
		Parameter & This work & \citet{naze2013} \\
		\hline
		$N_{\rm H}~[10^{22}$ cm$^{-2}]$    & $1.97 \pm 0.2$          & $1.1+ (0.4 \pm 0.08)$  \\
		$kT$ [keV]                     & $0.75^{+0.12}_{-0.11}$ & $0.93 \pm 0.07$      \\
		norm [$10^{-4}$ cm$^{-5}$]     &  $1.65^{+0.65}_{-0.42}$         &   $1.8 \pm 0.2$       \\
		$c_{\rm MOS1}$ & $0.99 \pm 0.2$ &  -  \\
		$c_{\rm MOS2}$ & $0.98 \pm 0.2$ &  -  \\
		Reduced $\chi^{2}$                     & 0.95                   & 1.12          \\ 
		\hline
	\end{tabular}
\end{table}

Then, we calculated the quotient $L_{\rm X}/L_{\rm BOL}$, which has a typical value of $10^{-7}$ for the winds of the O-type stars (\citealt{naze2013,gomezmoran2018} and references therein). It is assumed that for larger values, the emission comes from the colliding winds region, otherwise it comes from the winds of the components \citep{sana2006,chlebowski1984,chlebowski1989}.
To get this quotient, we took the bolometric luminosities considered in Sect.~\ref{sec:analysis} and estimated the X-ray luminosity of the system by calculating the flux $F_{\rm X}$ from the fitted model of the spectrum. We used the {\sc flux} task of {\sc xspec}, which represents the flux in X-rays corrected by ISM absorption, in the range of energy where HM1~8 emits: 0.5-3~keV. From this, we obtained
$F_{\rm X}=2.31^{+0.11}_{-0.27} \times 10^{-14} \text{ergs}^{-1}~\text{cm}^{-2}$.

Then, taking into account the distance we found the X-ray luminosity 
$L_{\rm X} = 2.2^{+0.3}_{-0.5}\times 10^{31}~\mathrm{erg~s}^{-1}$
Finally, we obtained the quotient $L_{\rm X}/L_{\rm BOL} \approx (0.19^{+0.03}_{-0.04}) \times 10^{-7}$,
which would indicate that the radiation comes from the stellar wind of the primary component.
This is different of what \citet[][]{naze2013} found, probably due to they considered the X-ray flux up to 8~keV and in this work we cut it at 3~keV because HM1~8 has not detectable emission beyond this point.

\begin{figure}
\includegraphics[width=\columnwidth]{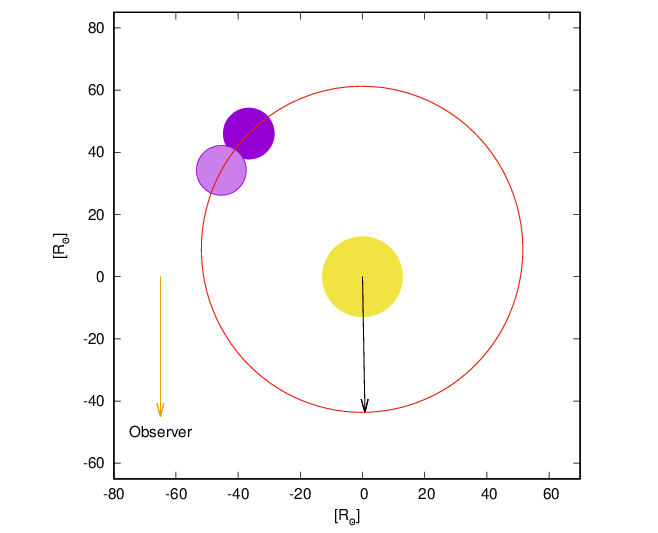}
    \caption{Configuration of the HM1~8 binary system during the X-ray observation of March 10, 2010.
    Yellow filled circle: primary component.
    Dark (light) purple filled circle: secondary component position at the start (end) of X-ray observation.
    Red ellipse: relative orbit of the secondary component.
    Orange arrow: direction to the observer.
    Black arrow: periastron position of the relative orbit.}
    \label{fig:orbfinal}
\end{figure}

One may wonder if a colliding winds region actually exists but was not visible at the time of the X-ray observation so, to discard it, we made a scheme of the stars positions during the {\it XMM-Newton} observation. As can be seen in Figure~\ref{fig:orbfinal}, if a colliding winds region exists, it should be detected since this region shall be very close to the surface of the secondary component, which was visible during the observation. 
Moreover, we do not see in the optical spectra indications of a possible colliding winds influence in the $H_{\alpha}$ line since it is in absorption in all the spectra.

\section{Tidal Evolution of the Pair} 
\label{Sec:tidal}

We now study the tidal evolution of HM1~8 with the intention of comparing the results of the models with the observed state of the system. 
In particular, since an orbital eccentricity was found from the observational analysis\footnote{This is a clear indication that the system has not suffered a Roche Lobe overflow (RLOF) yet and the pair is in a detached configuration. If the system were in the RLOF state, the orbit would be almost circular}, it is interesting to compute the circularisation timescale and compare it with the age of the system.

In order to compute the tidal evolution of the pair we need to solve a system of six ordinary, non linear differential equations. These correspond to the evolution of the major semiaxis of the orbit $a$ (or, equivalently, the orbital period $P_{\rm orb}$), the eccentricity $e$, the angular rotation $\omega_{\rm i}$ of each component and the inclination of their rotational axes $i_{\rm i}$ with respect to the orbital plane.

The tidal equations we solve are given by \citet{repetto2014} that are a generalization of those given by \citet{1981A&A....99..126H} (see also \citealt{2008ApJS..174..223B}).  These equations have been solved by a fully implicit, finite differences algorithm suitable for problems that may reach an equilibrium situation.

We assumed that the initial masses of the components
were of $33.8{\rm M}_{\odot}$ and $17.9{\rm M}_{\odot}$. If we assume an age of 2~Ma, these masses on the ZAMS evolve to the observed values. As stated above, we assumed solar composition, moderate overshooting and neglected the effects of rotation on the evolution of the components. Nevertheless, stellar rotation is considered in the tidal evolution.

This treatment cannot be considered as contradictory: Rotation of the components is far slower than critical rotation (and so, internal mixing due to rotation has a minor effect). On the contrary, considering the exchange between the angular momentum contained in the rotation of the components and the orbit is essential for a correct treatment of the tidal evolution.  

\begin{figure}
 \includegraphics[width=\columnwidth]{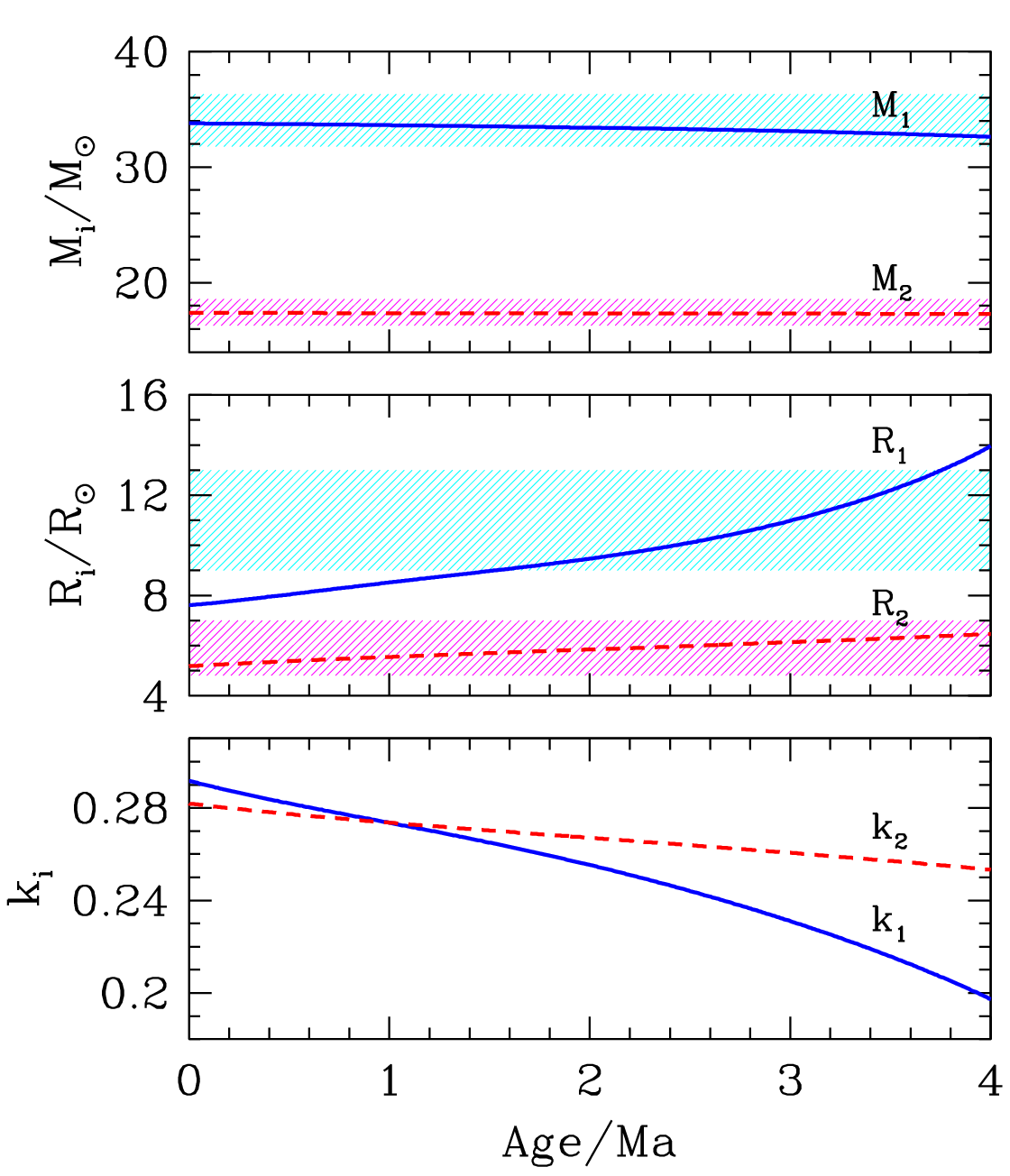}
 \caption{Evolution of the masses (upper panel), radii (middle panel) with their corresponding uncertainties. In the lower panel we present the radii of gyration for the components of HM1~8.}
 \label{Fig:estrellas}
\end{figure}

In order to solve the tidal evolution equations we need to know the evolution of some important quantities of each component of the pair. These are the masses, radii, and radii of gyration $k$ of the components ($I= k^{2} M R^{2}$, where $I$ is the moment of inertia of the star). They have been computed with the stellar code presented in \cite{2003MNRAS.342...50B} and the results are presented in Fig.~\ref{Fig:estrellas}. Since these are smooth functions, we tabulated them in advance, neglecting the effects of tides on the  evolution of this pair. This is a natural approximation, since  HM1~8 is still a well detached pair.

To explore the tidal evolution of the pair, we shall consider three different situations by assuming different initial conditions. Cases~I, II, and III are defined in order to analyse the sensitivity of the tidal evolution of the binary system under variations of the initial orbital period, eccentricity and angular velocity of rotation of the primary. We also tested the variation of the tidal evolution by changing the inclination of the axis of rotation of the components, as well as by changing the angular velocity of the secondary component. As these variations have a small impact on the tidal evolution of the pair we shall restrict ourselves to the discussion of the results corresponding to cases~I, II, and III.

We define $\varpi_{i}=(\omega_{i})_{\rm init}/(\omega_{i})_{\rm cr}$, where $(\omega^{2}_{i})_{\rm cr}= G M_{i}/R^{3}_{i}$ is the critical rotation rate if the star is spherical. It may be adequate to remark that here we employ $(\omega_{i})_{\rm cr}$ only to gauge the relevance of rotation on the structure. Considering that the components of  HM1~8 have projected tangential velocities of rotation of the order of 100~km~s$^{-1}$ and the values presented in Fig.~\ref{Fig:estrellas} for the masses and radii, we find that $\varpi_{i}\approx 1/7$ which, as discussed above, justifies the employment of evolutionary tracks of non rotating models.

\begin{itemize}

 \item Case~I: Consider $(e)_{\rm init}=0.29$; 
 $\varpi_{1}=0.20$; $(i_{1})_{\rm init}= 1.0$;
 $\varpi_{2}=0.20$; $(i_{2})_{\rm init}= 1.0$ and vary the orbital period:  $(P_{\rm orb})_{\rm init}=$ 5.705, 5.848, 5.994, 6.144, and 6.297~days. 
 
 \item Case~II: Consider $(P_{\rm orb})_{\rm init}=$ 6.000~days; $\varpi_{1}=0.20$; $(i_{1})_{\rm init}= 1.0$;
 $\varpi_{2}=0.20$; $(i_{2})_{\rm init}= 1.0$  and vary the eccentricity  $(e)_{\rm init}=$ 0.20, 0.220, 0.242, 0.266, and 0.293.
 
 \item Case~III: Consider $(P_{\rm orb})_{\rm init}=$ 5.900~days; $(e)_{\rm init}=0.20$;  $(i_{1})_{\rm init}= 1.0$;  $\varpi_{2}=0.20$; $(i_{2})_{\rm init}= 1.0$  and vary the angular velocity of the primary
 $\varpi_{1}=$ 0.100, 0.150, 0.225, 0.337, and 0.506.

\end{itemize}

\begin{figure}
 \includegraphics[width=\columnwidth]{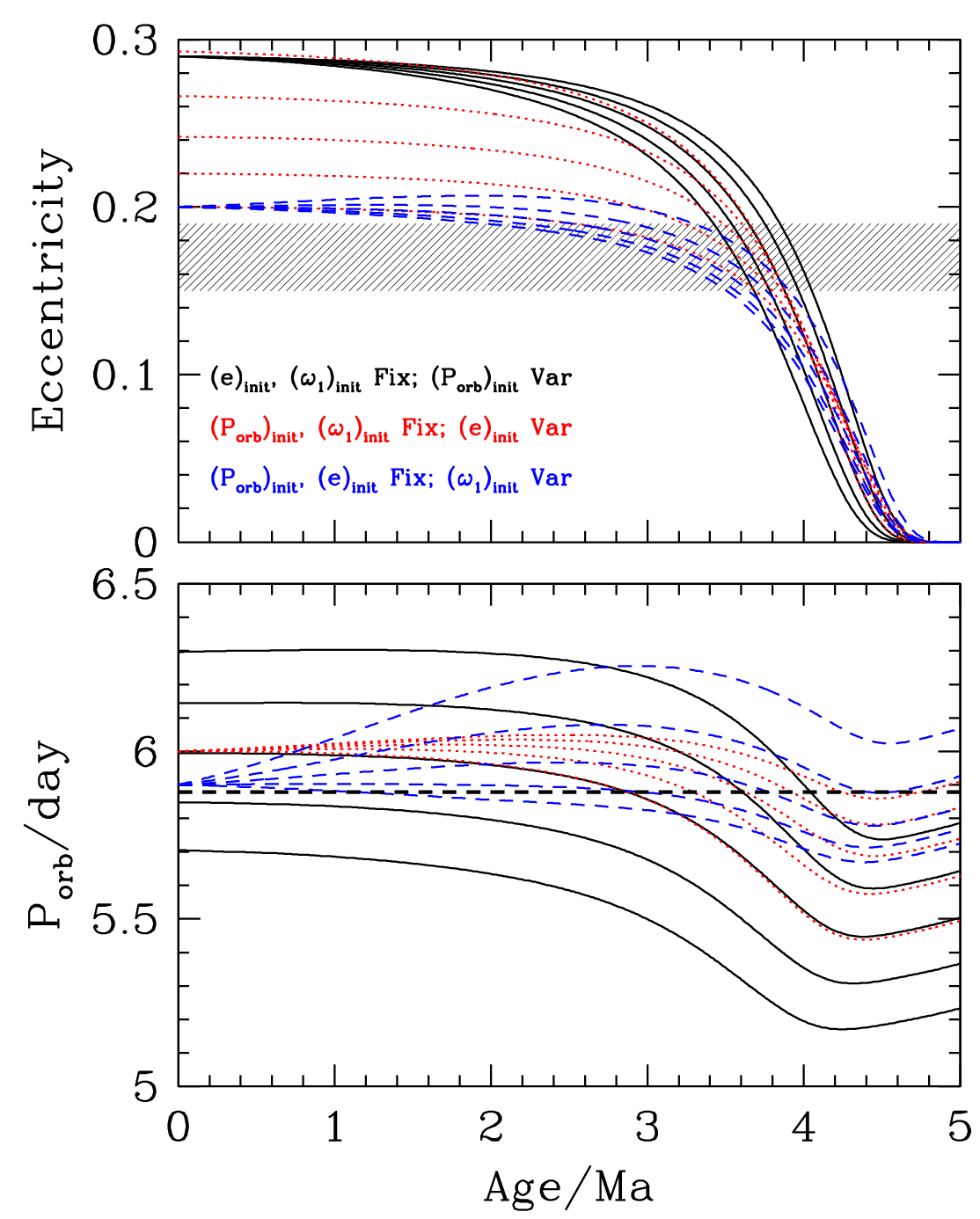}
 \caption{Tidal evolution of the HM1 8 system according to the models described in this paper. The upper and lower panels depict the time evolution of the system eccentricity and period, respectively. Black solid curves represent case~I: time evolution  for  5 different values of the initial orbital period. The upper curve corresponds to the larger $P_{\rm orb,init}$ for which the orbital period and the eccentricity show the slower evolution with time. Red dotted lines represent Case~II: time evolution for 5 different values of the initial eccentricity. The upper curve corresponds to the larger $(e)_{\rm init}$ for which the changes in e and P with evolution are also slower. Blue dashed lines represent Case~III: time evolution of the system for 5 different values of the primary rotational velocity. Again, the upper curve corresponds to the larger $\varpi_{\rm 1,init}$ for which the period and eccentricity show the slower evolution. In the upper panel, tilted dashed lines indicate the range of eccentricities compatible with observations. In the lower panel, the period is denoted with a horizontal dashed line.}
 \label{Fig:tidales}
\end{figure}

The main results of these simulations are shown in Fig.~\ref{Fig:tidales}. From these results we find that if the initial eccentricity of the pair was larger than the one now observed, there is a clear tendency of the system to circularisation and the timescale to reach the observed value is of the order of two to four million years, comparable to the evolutionary timescale of the primary component of the pair (see upper panel of Fig.~\ref{Fig:tidales}). This also tell us that, as the system is not yet circularised, consistent with an age less than 4~Ma. In the lower panel of Fig.~\ref{Fig:tidales} we show the evolution of the orbital period of the pair. It changes on a narrow interval, since the binary components are at a relative distance appreciably larger than their respective sizes. As a result, the exchange of angular momentum between components is rather weak.

\begin{figure}
 \includegraphics[width=\columnwidth]{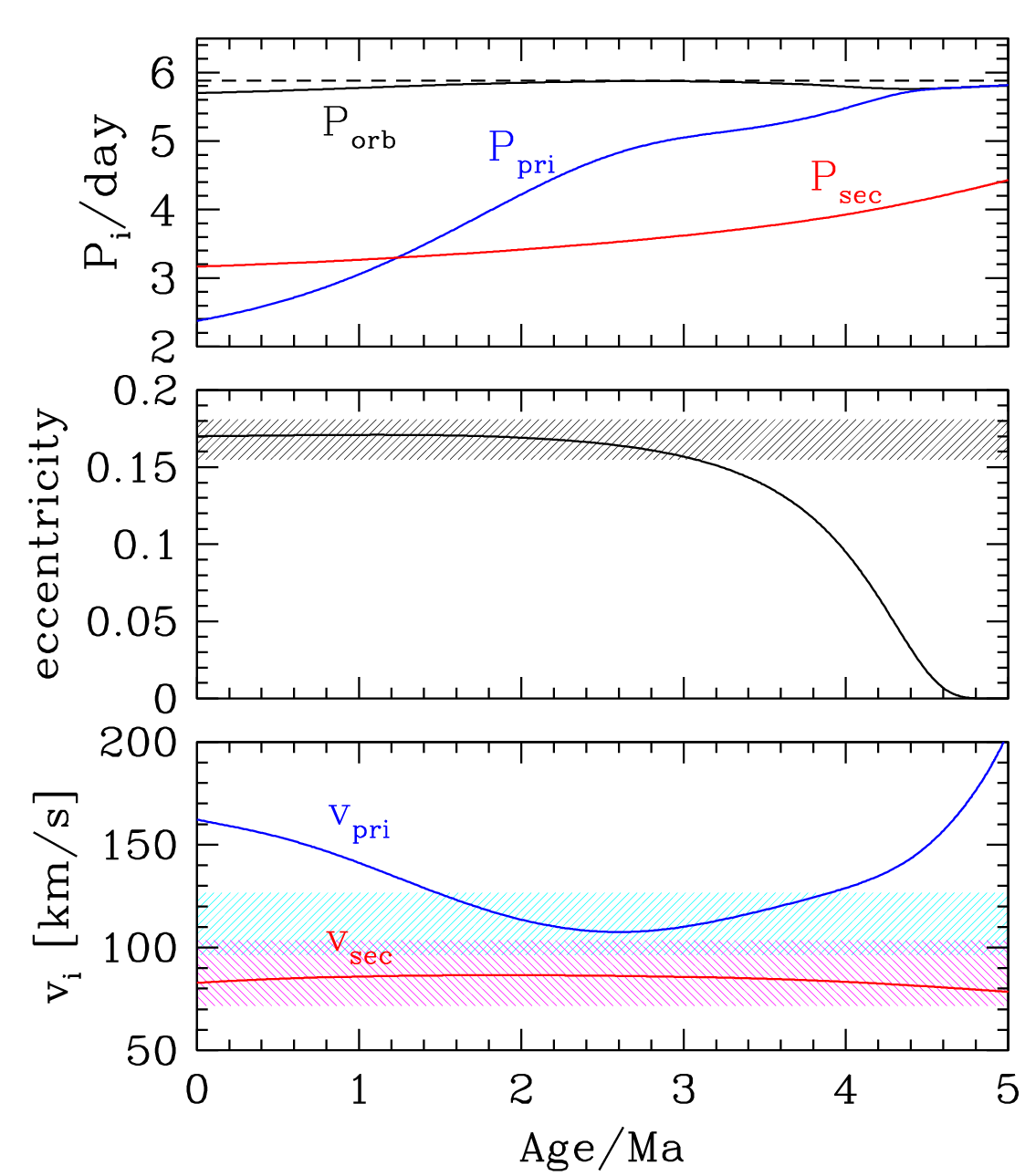}
 \caption{Evolution of the periods of the orbit and of rotation of each component, eccentricity of the orbit, and rotational velocities of a system with the masses corresponding to the case of HM1~8. Here we assumed $(P_{\rm orb})_{\rm init}= 5.70$~days,$(P_{\rm pri})_{\rm init}= 2.37$~days,$(P_{\rm sec})_{\rm init}= 3.17$~days, $(e)_{\rm init}=0.17$;  $(i_{1})_{\rm init}= (i_{2})_{\rm init}= 0.0$. For this particular configuration we find that at the age of 2~Ma this configuration is in agreement with the orbital period (represented with a dashed line in the upper panel), eccentricity and rotational velocities. In the lower panel the cyan (magenta) shaded region represents the uncertainty in the rotational velocity for the primary (secondary) component of the pair. Notice that the primary is synchronized to the orbit in $\approx 4$~Ma while the secondary does not reach this state in 5~Ma of evolution.}
 \label{Fig:tidales2}
\end{figure}

For completeness, in Fig.~\ref{Fig:tidales2} we present the tidal evolution of a system with the masses observed in HM1~8 that at an age of 2~Ma is in agreement with of observations. This corresponds for a particular set of initial conditions. In fact there should be a set of infinitely degenerate initial conditions that would also give compatibility with observations. While it is not the goal of this section to perform an exploration of these conditions, the chosen example proves that it is possible to get a nice agreement between the standard tidal evolution theory and our observations.

\section{Summary and conclusions} 
\label{sec:conclusions}

Using optical photometric and spectroscopic data, and X-ray observations, a comprehensive study of the massive binary system HM1~8 was performed. This analysis includes the computation of an improved orbital solution, updated spectral classifications of both stellar components, and the determination of the fundamental parameters of the system. 
The tidal evolution of the binary was also modelled. 

A spectral disentangling method applied to the observed composite spectra lead to the individual spectrum of each component, over which RV and quantitative spectral analyses were performed.  

The binary star HM1~8 is a very massive eclipsing system whose components, classified as O4.5~IV and O9.7~V, are characterized by a period of P=5.8782~days. 

The orbital period we obtained is very close to the value reported by \citet{gamen2008}.  
We were able to fit a spectroscopic orbital solution that indicates an eccentric orbit (e=0.14) with its major axis pointing very close to our line of sight. 
The fitted mass ratio $q \sim 0.5$ is the most common value in massive binary systems according to \citet{barba2017}. 

We detected a decrease in brightness synchronised with the phase of conjunction, according to the spectroscopic orbital solution, which corresponds to a partial eclipse of the primary star from the secondary one and allowed us to obtain the inclination of the system (i=70$^\circ$) and, with it, the absolute masses of the components (M$_a$=33.6~M$_{\sun}$ and M$_b$=17.7~M$_{\sun}$).

It is common knowledge the importance of determining reliable masses, and how poorly known they are for early-type stars, therein lies the relevance of the determination of a Galactic O4.5~IV mass by the method of eclipsing binaries. Before this work, the earliest giant star with a known mass (also obtained by the eclipsing binaries method) was HD152248b, an O7~III star with $M = 30.1\pm 0.4$~M$_{\sun}$ \citep{mayer2008}, and the earliest subgiant was HD152218a, an O9~IV, with $M = 19.8\pm 1.5$~M$_{\sun}$ \citep[][]{rauw2016}. 
For the secondary, the dynamical mass we obtained is $M_2 \approx 18$~M$_{\sun}$; and the other Galactic O9.7~V star mass, calculated with the eclipsing binary method, is HD152218b \citep[][]{rauw2016} with $M=15.0\pm1.1$~M$_{\sun}$.

We also performed a new X-ray spectroscopic analysis of the {\it XMM-Newton} data and obtained plasma parameters similar to those from \citet{naze2013}. Comparing the X-ray flux with the bolometric luminosity we estimated that the X-ray emission of HM1~8 originates in the primary wind and that there is not a colliding winds region. 

We studied the tidal evolution of the binary system by solving the equations of \citet{repetto2014}. As circularisation occurs on a timescale of 4~Ma and the expected age of the system is shorter, this provides a description consistent with the eccentric solution of the orbit deduced from observations. Furthermore, the evolution of the period shows slight changes along time, which correspond to a weak exchange of angular momentum, as expected for a binary system in which the distance between the components is appreciably larger than their sizes. 

At last, we shall give some context to this study of the HM1~8 system. First, it is important though not very common to obtain absolute dynamical masses of stars, in general due to the lack of information about orbital inclinations. In this case, it takes another level of importance since we have two massive components; furthermore, one of them is an O4.5 IV star which is the earliest subgiant star with a known dynamical mass.
Finally, it is notable that all the different techniques we used to analyse this system gave us consistent results; therefore we have a comprehensive view of the fundamental parameters, the behaviour and the evolution of this massive binary system.

\section*{Acknowledgements}

The authors warmly thank the referee, Ian Howarth, for his detailed review, suggestions and guidance that helped improving the manuscript. We thank the directors and staff at LCO, CTIO and La Silla for the use of their facilities and their kind hospitality during the observing runs.
C.N.R. acknowledges support from Consejo Nacional de Investigaciones Cient\'ificas y T\'ecnicas (CONICET) through the Beca Interna Doctoral grant and from Asociaci\'on Argentina de Astronom\'ia through the Beca de Est\'imulo a las Vocaciones Cient\'ificas grant.
C.N.R., G.A.F., and R.G. acknowledge support from grant PICT 2019-0344.
R.H.B. acknowledges support from ANID FONDECYT Regular Project No. 1211903.
We thank Sergio Sim\'on-D\'iaz for his valuable help with the {\sc iacob} tools.
This work was based on observations obtained with {\it XMM-Newton}, an ESA science mission with instruments and contributions directly funded by ESA Member States and NASA. 
This research has made use of the NASA's Astrophysics Data System and the SIMBAD database, operated at CDS, Strasbourg, France.

\section*{Data Availability}

The optical spectroscopic data underlying this article belongs to the {\it OWN Survey} team and will be shared on reasonable request to Dr. Rodolfo H. Barba; while the optical photometric data are available in the article and in its online supplementary material. Finally, the X-ray data underlying this article are available in the {\it XMM-Newton} Science Archive at \url{http://nxsa.esac.esa.int/nxsa-web/#search}, and can be accessed with the observation ID 0600080101.



\bibliographystyle{mnras}
\bibliography{biblio}



\bsp	
\label{lastpage}
\end{document}